\documentclass[longauth]{aa}  

\usepackage{graphicx}
\usepackage{chemformula}
\usepackage{txfonts}
\usepackage{caption}
\usepackage{placeins}
\usepackage{amsmath}    
\usepackage{amssymb}
\usepackage{CJKutf8}
\usepackage{threeparttablex}
\usepackage{hyperref}
\hypersetup{
  colorlinks   = true, 
  urlcolor     = blue, 
  linkcolor    = blue, 
  citecolor   = blue 
}
\usepackage{natbib}
\usepackage{orcidlink}
\bibpunct{(}{)}{;}{a}{}{,}

\begin{document}

   \title{TOI-2147\,b and TOI-6019\,b: Two eccentric warm Jupiters detected and characterized with TESS and MaHPS}
   
   \author{Luis Thomas\thanks{Email:lthomas@mpe.mpg.de}\orcidlink{0009-0006-1571-0306}\inst{1,2}
            \and
          Louise D. Nielsen \orcidlink{0000-0002-5254-2499} \inst{1}
          \and
          Hanna Kellermann \orcidlink{0009-0006-3527-0424} \inst{1,2}
          \and
          Bibiana Prinoth \orcidlink{0000-0001-7216-4846} \inst{3,4} \thanks{ESO Fellow}
          \and
          Yutong Liu \inst{1}
          \and
          Elif Zeynep Özden \inst{1}
          \and
          Arno Riffeser \inst{1}
          \and
          Claus Gössl \orcidlink{0000-0002-2152-6277}\inst{1}
          \and
          Frank Grupp \inst{1,2}
          \and
          Jerome de Leon \orcidlink{0000-0002-6424-3410} \inst{5}
          \and
          Karen A. Collins \orcidlink{0000-0001-6588-9574} \inst{6}
          \and
          Allyson Bieryla \orcidlink{0000-0001-6637-5401} \inst{6}
          \and
          Lorena Acuña-Aguirre \orcidlink{0000-0002-9147-7925} \inst{7}
          \and
          Keith Baka \orcidlink{0009-0003-4963-3255} \inst{1} 
          \and
          Malte Busmann \orcidlink{0009-0001-0574-2332} \inst{1,8} 
          \and
          David R. Ciardi \orcidlink{0000-0002-5741-3047} \inst{9}
          \and
          Catherine A. Clark \orcidlink{0000-0002-2361-5812}  \inst{9}
          \and
          Juliana Ehrhardt \orcidlink{0009-0003-9433-043X} \inst{1,2}
          \and
          Mark E. Everett \orcidlink{0000-0002-0885-7215} \inst{10}
          \and
          Akihiko Fukui \orcidlink{0000-0002-4909-5763} \inst{5}
          \and
          Jan-Vincent Harre \orcidlink{x0000-0001-8935-2472} \inst{1} 
          \and
          Keisuke Isogai \orcidlink{0000-0002-6480-3799} \inst{11,12}
          \and
          Yanxi Li (\begin{CJK}{UTF8}{gbsn}李言蹊\end{CJK}) \orcidlink{0009-0001-8212-102X} \inst{13}
          \and
          Felipe Murgas \orcidlink{0000-0001-9087-1245} \inst{14,15}
          \and
          Norio Narita \orcidlink{0000-0001-8511-2981} \inst{5,16,14}
          \and
          Enric Palle \orcidlink{0000-0003-0987-1593} \inst{14,15}
          \and
          Hannu Parviainen \orcidlink{0000-0001-5519-1391} \inst{14,15}
          \and
          Jan-Niklas Pippert \orcidlink{0009-0006-9461-002X} \inst{1,2}
          \and
          Christoph Ries \inst{1}
          \and
          Boris S. Safonov \orcidlink{0000-0003-1713-3208} \inst{17}
          \and
          Thomas Schäfer \inst{18}
          \and
          Michael Schmidt \inst{1}
          \and
          Richard P. Schwarz \orcidlink{0000-0001-8227-1020} \inst{6}
          \and
          Laura Schöller \orcidlink{0009-0008-2469-0372} \inst{2}
          \and
          Gregorg Srdoc \inst{19}
          \and
          Ivan A. Strakhov \orcidlink{0000-0003-0647-6133} \inst{17}
          \and
          Suzanne Taylor \orcidlink{0009-0004-7966-2812} \inst{20}
          \and
          Wenjie Zhou (\begin{CJK}{UTF8}{gbsn}周文杰\end{CJK}) \orcidlink{0000-0002-8385-7194} \inst{13}
          \and
          Raphael Zöller \orcidlink{0000-0002-0938-5686} \inst{1,2}
}

   \institute{
   University Observatory Munich, Faculty of Physics, Ludwig-Maximilians-Universit\"at München, Scheinerstr. 1, 81679 Munich, Germany
    \and
    Max-Planck Institute for Extraterrestrial Physics, Giessenbachstrasse 1, D-85748 Garching, Germany
    \and
    European Southern Observatory (ESO), Karl-Schwarzschild-Str. 2, 85748 Garching bei München, Germany
    \and
    Lund Observatory, Division of Astrophysics, Department of Physics, Lund University, Box 118, 221 00 Lund, Sweden 
    \and
    Komaba Institute for Science, The University of Tokyo, 3-8-1 Komaba,
    Meguro, Tokyo 153-8902, Japan
    \and
    Center for Astrophysics \textbar \ Harvard \& Smithsonian, 60 Garden Street, Cambridge, MA 02138, USA
    \and 
    Max Planck Institut für Astronomie, Königstuhl 17, 69117 Heidelberg, Germany
    \and
    Excellence Cluster ORIGINS, Boltzmannstr. 2, 85748 Garching, Germany
    \and
    NASA Exoplanet Science Institute - Caltech/IPAC, Pasadena, CA 91125 USA
    \and
    NSF NOIRLab, 950 N. Cherry Ave., Tucson, AZ 85719, USA
    \and
    Okayama Observatory, Kyoto University, 3037-5 Honjo, Kamogatacho,
    Asakuchi, Okayama 719-0232, Japan
    \and
    Department of Multi-Disciplinary Sciences, Graduate School of Arts and
    Sciences, The University of Tokyo, 3-8-1 Komaba, Meguro, Tokyo
    \and
    Xingming Observatory, Urumqi, Xinjiang, China
    \and
    Instituto de Astrof\'{i}sica de Canarias (IAC), 38205 La Laguna, Tenerife, Spain
    \and
    Departamento de Astrof\'isica, Universidad de La Laguna (ULL), E-38206 La Laguna, Tenerife, Spain
    \and
    Astrobiology Center, 2-21-1 Osawa, Mitaka, Tokyo 181-8588, Japan
    \and
    Sternberg Astronomical Institute Lomonosov Moscow State University, Universitetskii prospekt, 13, Moscow, Russia
    \and
    Citizen Scientist, Zooniverse c/o University of Oxford, Keble Road, Oxford OX1 3RH, UK
    \and
    Kotizarovci Observatory, Sarsoni 90, 51216 Viskovo, Croatia
    \and
    Western Colorado University, 1 Western Way, Gunnison, CO 81230, USA
    }
   \date{}

 \titlerunning{TOI-2147\,b and TOI 6019\,b: Two eccentric WJs}
  \abstract
{The population of Jupiter-sized exoplanets with orbital periods between 10 and 200 days (WJs) exhibits a broad range of orbital eccentricities and system architectures, suggesting a diversity of formation and migration pathways. In this work, we report the detection and characterization of two new eccentric WJs, TOI-2147\,b and TOI-6019\,b, initially identified as planet candidates by the Transiting Exoplanet Survey Satellite (\textit{TESS}). We combined TESS photometry with ground-based follow-up observations, including multiband photometry from LCOGT and MuSCAT2, high-angular-resolution speckle imaging, and high-precision radial velocity measurements from the high-resolution Manfred Hirt Planet Finder Spectrograph (MaHPS). Using these data, we were able to confirm the planetary nature of both candidates. TOI-2147\,b has a radius of $10.5 \pm 0.3\,\mathrm{R}_\oplus$ and a mass of $116 \pm 22\,\mathrm{M}_\oplus$. It orbits its slightly metal-poor ($\mathrm{[Fe/H]} = -0.29^{+0.07}_{-0.08}$) G-type host star on an eccentric orbit ($e = 0.29 \pm 0.07$) with a period of 26.2 days. TOI-6019\,b has a radius of $12.3 \pm 0.3\,\mathrm{R}_\oplus$ and a mass of $149 \pm 15\,\mathrm{M}_\oplus$. It orbits a slightly evolved, solar-metallicity G-type sub-giant with a period of 14.5 days on a significantly eccentric orbit ($e = 0.48^{+0.05}_{-0.04}$). Both planets have bulk densities below that of Jupiter, indicating mildly inflated radii, with interior structure modeling using \texttt{GASTLI}. This suggests that tidal heating from the nonzero eccentricities likely contributes to this inflation and disfavors large atmospheric metal enrichment. No significant signals from additional companions were detected in the radial velocity time series or transit timing variations. Together with the elevated eccentricities, this is consistent with a high-eccentricity migration origin for both systems. }


   \maketitle
%

\section{Introduction}
The discovery of giant exoplanets orbiting close to their host stars has revolutionized our understanding of the formation and evolution of planetary systems. Hot Jupiters (HJs), with orbital periods of fewer than 10 days, and warm Jupiters (WJs), with periods between 10 and 200 days, challenge classical planet formation theories, which predict that gas giants should form beyond the ice line at orbital distances of several astronomical units \citep{Pollack1996,Lecar2006,Kennedy2008}. The mechanisms responsible for their observed location remain a subject of debate, with leading hypotheses including high-eccentricity tidal migration, disk-driven migration, and in situ formation \citep[a summary is provided in][]{Dawson2018}.

While in situ formation might be feasible under special conditions \citep{Batygin2016,Boley2016}, the prevalent explanation is that giant planets form beyond the ice line and migrate inward through high-eccentricity or disk-driven migration. Disk-driven migration posits that giant planets migrate inward through interactions with the protoplanetary disk \citep{Lin1986,Goldreich1980,Lin1996}. For high-eccentricity migration, the giant planets are placed on eccentric orbits through dynamical interactions such as planet-planet scattering \citep{Rasio1996,Chatterjee2008}, Lidov–Kozai cycling \citep{Wu2003,Fabrycky2007,Naoz2016,Petrovich2016,Vick2019}, and secular chaos \citep{Wu2011,Petrovich2015,Teyssandier2019}. The orbit then circularizes and shrinks through tidal interaction with the star at periastron passage, delivering the planet to its present-day orbit. 

The system architecture and properties of warm and HJs can give information on their dynamical history and possible migration mechanisms. Most HJs have no close-companion planets that would plausibly be destroyed during high-eccentricity migration \citep{Mustill2015,Sha2026}, but show signs of massive long-period companions, which could instigate the migration \citep{Knutson2014,Zink2023}. Furthermore, HJs might be more abundant in dense stellar environments where increased dynamical interaction might lead to a higher migration rate \citep{Brucalassi2017,Thomas2024}. These results point toward high-eccentricity migration as the dominant formation mechanism for HJs. However, some HJs have been found to possess smaller close-in companions \citep[e.g., ][]{Becker2015,Korth2023,Korth2024,Grieves2025}. Additionally, analysis of transit timing variations found that $12\pm 6\%$ of HJs might host nearby companions \citep{Wu2023}, indicating a more diverse formation and migration pathway. On the other hand, WJs are found to exist preferentially in multiplanet systems with nearby companions, pointing to more quiescent formation mechanisms \citep{Wu2023}. Measurements of the sky-projected stellar obliquity through the Rossiter-McLaughlin (RM) effect support these results. HJs orbiting cooler stars below the Kraft break are preferentially aligned, while planets around hot stars have a diverse obliquity distribution \citep{Winn2010,Albrecht2022}. These observations are consistent with high-eccentricity migration and the obliquity distribution being shaped by dissipative mechanisms \citep{Rice2022hot}. Overall, WJs are preferentially found in aligned orbits \citep{rice2022warm}, supporting quiescent migration.

The observed eccentricity distribution of WJs, however, paints a more diverse picture for their formation and migration, with a significant fraction ($\sim 70\%$) of WJs exhibiting dynamically "hot" orbits (e > 0.1)  \citep{Dong20212,Morgan2026}. Some of the highly eccentric WJs, such as TOI-3362~b \citep{Dong2021} and TIC~241249530~b \citep{Gupta2024}, are expected to end up as HJs after circularization, and thus they can be considered proto-HJs. 
The population of eccentric WJs could come from stalled high-eccentricity migration, where the eccentricity was not excited to a degree high enough to bring the planet close to the star for the tidal forces to fully circularize the orbit. Alternatively, they could migrate through disk migration and then have their eccentricities excited later in their lifetime. The eccentricities of WJs have been found to correlate with host star metallicity, with metal-rich stars harboring more eccentric planets \citep{Dawson2013,Morgan2026}. This correlation likely arises from the enhanced formation efficiency of giant planets in metal-rich disks, leading to increased dynamical interactions and eccentricity excitation \citep{Santos2004,Fischer2005,Johnson2010}.

In this paper, we present the detection and characterization of the two WJs TOI-2147\,b and TOI-6019\,b, initially reported as planet candidates by the Transiting Exoplanet Survey Satellite (\textit{TESS}). We obtained high-resolution spectroscopy observations using the Manfred Hirt Planet Finder Spectrograph (MaHPS) to confirm their planetary nature and measure precise masses. The structure of this paper is as follows. Section \ref{secobs} describes the observations used to detect and characterize the new WJs. Sections \ref{stellar} and \ref{Analysis} present the analysis and derivation of the stellar and planetary parameters. Section \ref{results_disc} discusses the properties of the two planets and places them within the population of known giant planets. Finally, Section \ref{sum} summarizes our conclusions.

\section{Observations} \label{secobs}

\subsection{TESS photometry}
\begin{figure*}
    \centering
    \includegraphics[width=1.9\columnwidth]{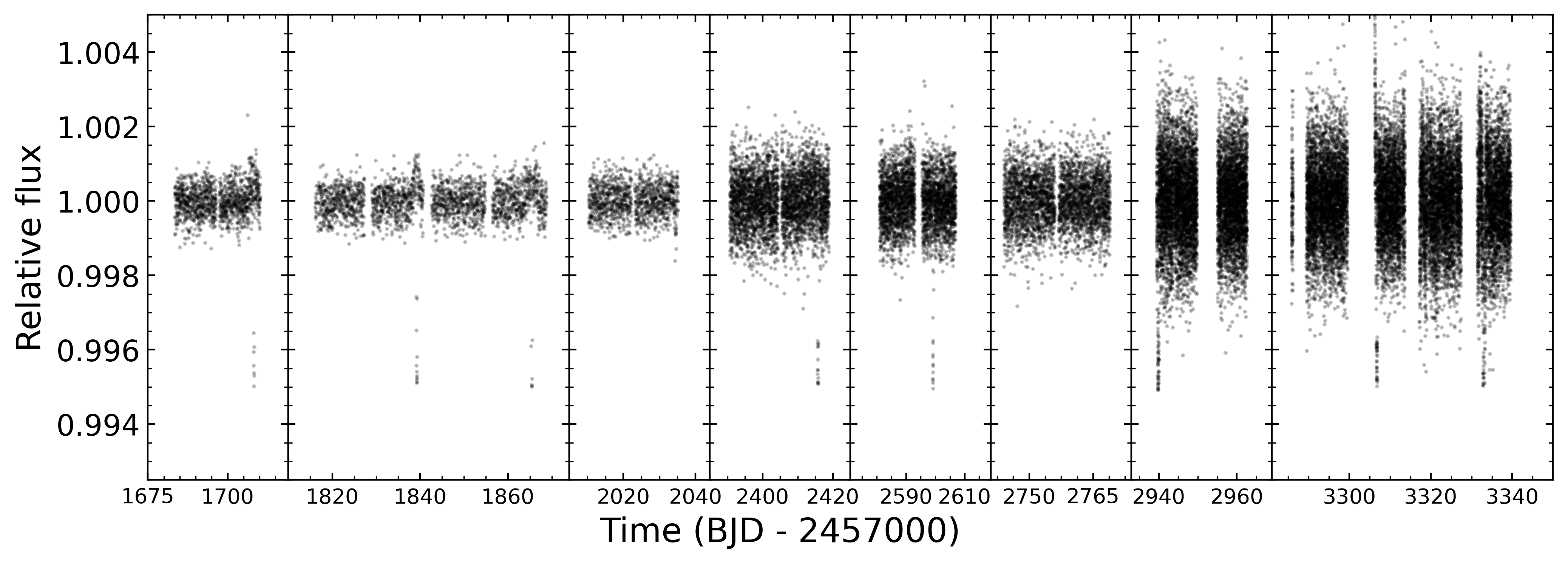} 
    \includegraphics[width=1.9\columnwidth]{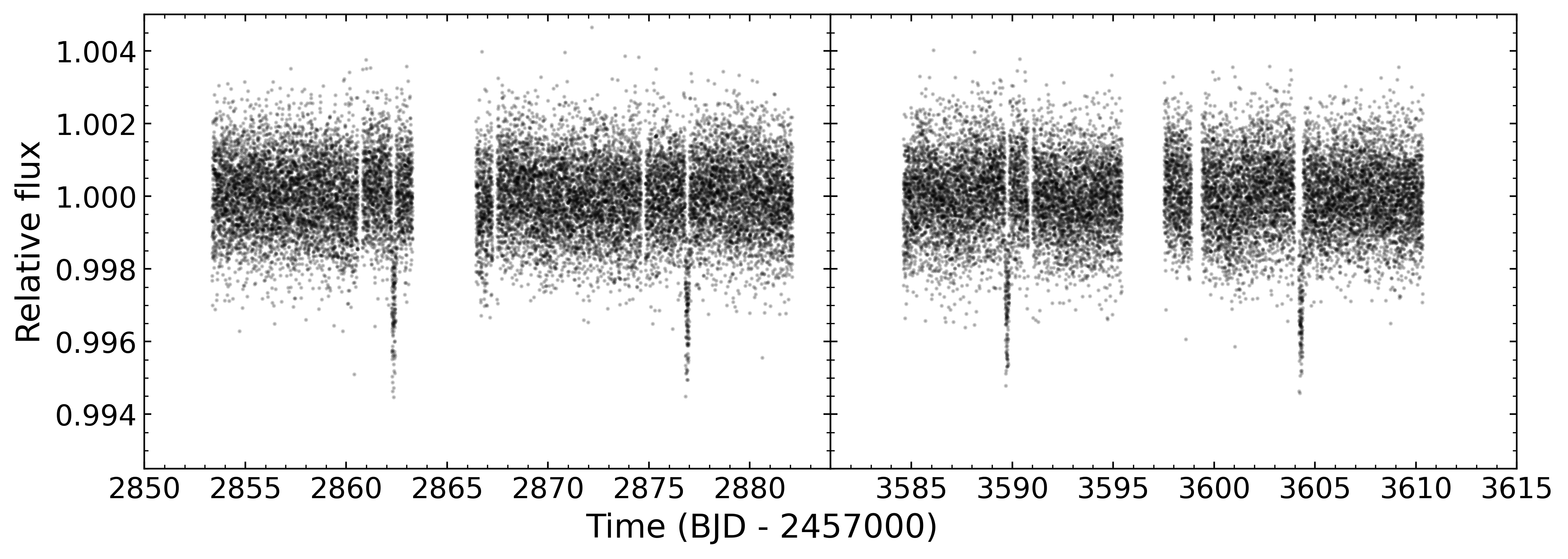} 
    \caption{PDCSAP from the \textit{TESS} light curves for TOI-2147 (upper plot) and TOI-6019 (lower plot). The increased scatter for the TOI-2147 in later sectors is due to the shorter exposure times.}
    \label{fig:tesslc}
\end{figure*}

TOI-2147 was observed in eight \textit{TESS} sectors (14, 19, 20, 26, 40, 47, 53, 60, 73, 74) at different cadences (30 min, 10 min, and 2 min). TOI-6019 was observed in Sector 57 and 84 with a 2-minute cadence. Transit events were detected by the Quick Look Pipeline \citep[QLP, ][]{huang2020photometry,huang2020b,kunimoto2021,Kunimoto2022} and the \textit{TESS} Science Processing Operation Center pipeline \citep[SPOC][]{jenkins2016tess,Jenkins2020}. We used the \texttt{lightkurve} package \citep{2018ascl.soft12013L} to download the TESS-SPOC High-Level Science Product \citep{caldwell2020tess} light curves. For our analysis, we used the presearch data conditioning simple aperture photometry  \citep[PDCSAP,][]{smith2012kepler,stumpe2012kepler,stumpe2014multiscale} light curves, which correct the simple aperture photometry (SAP) for various instrumental and systematic effects. The reported orbital periods and depths of the transit events by the \textit{TESS} team on ExoFOP are 26.2~days and 5593~ppm for TOI-2147.01 and 14.55~days and 3223~ppm for TOI-6019.01. The light curves are shown in Figure \ref{fig:tesslc}.

\subsection{Ground-based photometric follow-up}
The \textit{TESS} pixel scale is $\sim 21\arcsec$ pixel$^{-1}$ and photometric apertures typically extend out to roughly 1 arcminute, generally causing multiple stars to blend in the \textit{TESS} photometric aperture. To determine the true source of the \textit{TESS} detection, we acquired ground-based time-series follow-up photometry of the fields around TOI-2147 and TOI-6019 as part of the \textit{TESS} Follow-up Observing Program \citep[TFOP;][]{collins:2019}\footnote{https://tess.mit.edu/followup}. The on-target follow-up light curves are also used to place constraints on the transit depth and the \textit{TESS} ephemeris. Multiband observations place limits on transit depth chromaticity. We used the {\tt TESS Transit Finder}, a customized version of the {\tt Tapir} software package \citep{Jensen:2013}, to schedule our transit observations.

\subsubsection{LCOGT\label{subsubsec:lcogt}}
We observed a transit egress window of TOI-2147.01 in alternating $B$ and Pan-STARRS $z_s$ filters on February 14, 2022, from the Las Cumbres Observatory Global Telescope \citep[LCOGT;][]{Brown:2013} 1.0\,m network node at Teide Observatory on the island of Tenerife (Teide). Exposure times were set at 30\,s for the $B$ filter and 45\,s for the $z_s$ filter. The 1\,m telescopes are equipped with a $4096\times4096$ SINISTRO camera having an image scale of $0\farcs389$ per pixel, resulting in a $26\arcmin\times26\arcmin$ field of view. The images were calibrated by the standard LCOGT {\tt BANZAI} pipeline \citep{McCully:2018}, and differential photometric data were extracted using {\tt AstroImageJ} \citep{Collins:2017}. We used circular photometric apertures with radius $4\farcs5$, which excluded all of the flux from the nearest known neighbor in the Gaia DR3 catalog that is bright enough to possibly be the source of the TESS TOI-2147 transit event (Gaia DR3 1146601236892702720), which is $24\farcs4$ east of TOI-2147. Transit egress with consistent depth was detected in the target star apertures in both filter bands, thus confirming an achromatic transit-like event on-target. The light curve data are included in the joint modeling analyses described in Section \ref{Analysis}.

\subsubsection{New Mexico Skies-T19}
We observed the egress of TOI-2147.01 on March 13, 2022, in the Exo filter using the remote 0.43\,m iTelescope located during the time of observation at New Mexico Skies\footnote{https://support.itelescope.net/support/solutions/articles/251589}. Imaging was done with an SBIG STX-16803 CCD camera, which has an array of $4096\times 4096$ pixels with a pixel size of 9 $\mu \mathrm{m}$ x 9 $\mu m$. Weather conditions were good with clear skies. During the imaging session, the wind speed reached peaks up to 16.1 km/h, and the seeing was measured between 1.3 and 2.5 FWHM arcseconds. For the images, an exposure time of 60\,s was used. After visual inspection and blinking of all images, 23 images have been rejected due to poor guiding and defocus. For the analysis of the observable transit egress, a collection of 118 images has been finally selected for light curve analysis and transit model calculation.  

\subsubsection{MuSCAT2}
We observed an egress of TOI-6019.01 on September 23rd, 2023, in $g'$,$r'$,$i'$,$z_s$ with the MuSCAT2 \citep{Narita2019} instrument mounted on the 1.52\,m TCS telescope at the Teide Observatory in the Canary Islands, Spain.
MuSCAT2 is capable of simultaneous imaging in four bands, each using a $1\mathrm{k}\times1\mathrm{k}$ CCD camera with a pixel scale of 0."44 pixel$^{-1}$ and a field of view (FOV) of $7.\arcmin4\times7.\arcmin4$.
To avoid contamination from the nearby star (TIC\,125520908; $\Delta$Tmag$=7$; $r=8\arcsec$), we observed in focus and set the exposure times at (1, 2, 2, 2) s for $g'$,$r'$,$i'$,$z_s$, respectively.

We used the MuSCAT2 pipeline\footnote{https://github.com/hpparvi/MuSCAT2\_transit\_pipeline}, described in \citet{Parviainen2020c}, to perform the data reduction and aperture photometry over a set of comparison stars and aperture sizes. The optimal light curves were determined through a global optimization of a model, consisting of five brightest comparison stars and uncontaminated aperture radii smaller than $5.6\arcsec$, while the transit and baseline variations are simultaneously modeled using a linear combination of covariates. As reported in TFOP-SG1, we detected an ~on-time $(R_p/R_*)^2\sim3.5$~ppt transit using an achromatic transit model.

\subsection{High angular-resolution imaging}
To further rule out false-positive scenarios, we collected high angular-resolution imaging observations for both of our targets. We used these to search for unidentified close companions which could be the source of the transit signal, or dilute the transit depth, leading to an underestimated radius for the planet \citep{ciardi2015}.

We observed TOI-2147 on UT 2022 April 21 using the NN-EXPLORE Exoplanet Stellar Speckle Imager \citep[NESSI;][]{Scott2018}, a speckle imaging instrument at the WIYN 3.5\,m telescope on Kitt Peak. NESSI was operated with only one wavelength channel this night using a filter centered on $\lambda_c = 832$~nm with a width of 40~nm. The observation consisted of a sequence of $9000 \times 40\,$ms exposures. NESSI's field of view was set by a $256\times256$ pixel sub-array readout, resulting in $4.6\arcsec\times4.6\arcsec$ images.  However, we confined the contrast curve measurements to within $1.2\arcsec$ from the target star to avoid the effects of decorrelation in speckle patterns at wider separations.  A nearby point source calibration star was also observed immediately following the observation of the TOI for the purpose of estimating the PSF at the time of the science observation.

These speckle data were reduced using a pipeline described by \cite{Howell2011}. Among the pipeline products is a reconstructed image of the field around TOI-2147. The reconstructed image was used to measure a contrast curve, setting the detection limits on any nearby point sources. No companion stars were detected brighter than the contrast limit.  Initially, the contrast limit increases rapidly with angular separation from TOI-2147, from ${\Delta}m=0$ at the diffraction limit of $0.06\arcsec$ to ${\Delta}m=3.7$ at $0.15\arcsec$. The contrast increases more gradually past $0.15\arcsec$, reaching ${\Delta}m=5$ by $1\arcsec$.

TOI-6019 was observed on December 24, 2022, with the Speckle Polarimeter (SPP), a facility instrument on the 2.5-m telescope at the Caucasian Observatory of the Sternberg Astronomical Institute (SAI) of Lomonosov Moscow State University. A low-noise CMOS detector, Hamamatsu ORCA--quest \citep{Strakhov2023} was used. TOI-2147 was observed on October 27, 2020, with a previous, EMCCD--based version of the instrument \citep{Safonov2017}. In both cases, the atmospheric dispersion compensator was active, which allowed using the $I_\mathrm{c}$ band. The respective angular resolution was $0.083^{\prime\prime}$. The pixel scale was 20.6 mas/pixel. The power spectrum of TOI-6019 was estimated from 5200 frames with 23 ms exposure time. For TOI-2147, 4000 frames with 60 ms exposure were accumulated. The long–exposure atmospheric seeing was $0.65^{\prime\prime}$ and $1.05^{\prime\prime}$, respectively, for TOI-6019 and TOI-2147.

For TOI-6019, we detected a stellar companion at a separation $977\pm7$~mas and a position angle PA=$285.5\pm0.5^{\circ}$, with a magnitude difference is $6.6 \pm 0.3$~mag. The detection limits are $\Delta I_\mathrm{c}=4.2$ and $7.2$~mag at distances $0.25$ and $1.0^{\prime\prime}$ from the star, respectively. For TOI-2147, we did not detect any stellar companions; the detection limits are $\Delta I_\mathrm{c}=4.5$ and $5.8$~mag at distances $0.25$ and $1.0^{\prime\prime}$ from the star, respectively.

\subsection{High-resolution spectroscopy}
\subsubsection{TRES}
We obtained reconnaissance spectra with the Tillinghast Reflector Echelle Spectrograph (TRES, \cite{furesz2008design}) on the 1.5 m Tillinghast Reflector telescope at the Fred Lawrence Whipple Observatory (FLWO) on Mount Hopkins in Arizona. TRES is a fiber-fed, optical spectrograph with a resolving power of R = 44,000 covering a wavelength range between 390 and 910 nm. TOI-2147 was observed twice: on November 19, 2020 and December 2, 2020. TOI-6019 was observed on December 24, 2022 and January 1, 2023. All the spectra were visually inspected to check for composite spectra and then a multi-order spectral analysis was performed, following procedures outlined in \cite{buchhave2010} and \cite{Quinn2012}. We found no signs of composite spectra or larger RV variations indicative of an eclipsing binary. The differences in the RVs obtained at opposite quadratures are $\sim 52$\,m/s and $\sim 95$\,m/s with errors $\sim 35$\,m/s and $\sim 22$\,m/s for TOI-2147 and TOI-6019, respectively.

\subsubsection{MaHPS}
We confirmed the planetary nature of the two candidates and measured their masses using the Manfred Hirt Planet finder Spectrograph \citep[MaHPS, ][]{Kellermann2015,Kellermann2016} installed at the 2.1~m telescope of the Wendelstein Observatory in the Bavarian Alps in Germany. This spectrograph consists of the stabilized, fiber-fed, high-resolution ($R \sim 60,000$), optical ($385 - 885$ nm), echelle spectrograph FOCES \citep{Pfeiffer1998}, and a Menlo Systems Astro Frequency Comb as a wavelength calibration source. 

TOI-2147 was observed between the 26th of January, 2023, and the 14th of December 2025, collecting a total of 136 spectra on 57 nights. For TOI-6019, we collected 93 spectra (38 nights) between the 14th of July 2024 and the 18th of November 2025. With V-band magnitudes of 11.1 and 10.8, both stars are close to the limit for precise RV work with MaHPS \citep{Ehrhardt2024}. Therefore, we planned to collect three 1800\,s exposures per night to reach the required RV precision, which was not always possible due to worsening weather conditions or observations with a higher priority. 

The data were reduced and the RVs were extracted using our two in-house pipelines \texttt{GAMSE} \citep{Wang2016} and \texttt{MARMOT} \citep{Kellermann2020}, following the procedure outlined in \cite{Thomas2025}. We used all available observations to create the templates that were then used to calculate the RV shift using the template matching algorithm implemented in \texttt{MARMOT}. The MaHPS RVs used in the following analysis are available in Tables 5 and 6 at the CDS.

\section{Stellar analysis} \label{stellar}
\subsection{Spectroscopic stellar parameters} \label{stellarspec}
We characterized the host stars of the two planet candidates by combining spectroscopic stellar parameters with the available broadband photometry. We used the Python code {\tt SpecMatch-Emp} \citep{yee2017} to extract $T_\mathrm{eff}$, $\log g$, and $\text{[Fe/H]}$ from the MaHPS spectra. The MaHPS templates of the stars that were used for the RV extraction are compared order-wise to a library of 404 spectra in the wavelength range between 500 nm and 640 nm. The five best-fitting library spectra are then combined through linear interpolation to derive the final stellar parameters. We found $T_\mathrm{eff}=5917 \pm 110$\,K, $\log g = 4.23 \pm 0.12$ and $\text{[Fe/H]}=-0.28 \pm 0.09$ for TOI-2147 and $T_\mathrm{eff}=5635 \pm 110$~K, and $\log g = 4.01 \pm 0.12$ and $\text{[Fe/H]}=0.02 \pm 0.09$ for TOI-6019. These values are compatible with the stellar parameters that were derived using the stellar parameter classification tool \citep[SPC; ][]{buchhave2012,buchhave2014} on the TRES reconnaissance spectra.  The spectroscopic parameters are used as priors in the following analysis of the stellar properties.

\subsection{Bulk stellar parameters} \label{stellarsed}
To derive the bulk stellar parameters, we performed a combination of spectral energy distribution (SED) and isochrone fitting. The SED fitting is done with the Python package ARIADNE \citep{ariadne2022}, which uses Bayesian model averaging to combine the results of fitting different stellar atmosphere models. We fit stellar atmosphere models from PHOENIX V2 \citep{husser2013}, BT-Settl \citep{Hauschildt1999,allard2012}, Kurucz \citep{Kurucz1993}, and Castelli \& Kurucz \citep{Castelli2003} to broadband photometric data from the filters TYCHO $B$, $V$, Johnson $B$, $V$, Gaia DR3 $G$, $RP$, and $BP$, 2MASS $J$$H$$K$, and ALL-WISE $W$1 and $W$2. We used Gaussian priors on $T_\mathrm{eff}$, $\log g$, and $\log \text{[Fe/H]}$ from our spectroscopic analysis. For the distance, $D$, we used a normally distributed prior around the Bailer-Jones distance estimate from Gaia EDR3 \citep{Bailer-Jones2021}. The stellar radius, $R_*$ and extinction, $\text{A}_v$ are given uniform priors ranging from 0.05 to 20 R$_\odot$ and 0 to the maximum line-of-sight extinction calculated from the updated SFD galactic
dust map \citep{Schlegel1998,Schlafly2011} through the dustmaps Python package \citep{Green2018}.

To derive the mass and age of the two stars, we fit the results of the SED analysis to interpolated MESA Isochrones and Stellar Tracks \citep[MIST,][]{Dotter2016,Choi2016}, using the {\tt isoclassify} package \citep{Huber2017}. The final stellar parameters are summarized in Table \ref{tab:star}. For the projected rotational velocity, we adopted the values from the TRES spectra, which are also consistent with the expected values from the {\tt SpecMatch-Emp} analysis of the MaHPS spectra. With this data, we were able to calculate the maximum rotation period of the two stars using P$_{rot}/\sin i = \frac{2 \pi R_*}{v\sin i}$. 
\begin{table}
    \centering
    \caption{\label{tab:star}Stellar parameters of TOI 2147 and TOI 6019.}
    \resizebox{\columnwidth}{!}{
    \begin{threeparttable}[b]
    \centering
    
    \begin{tabular}{lccl}
    \hline
    \hline
      Parameters   & TOI 2147 & TOI 6019 & Ref. \\
      \hline
      \\
      Identifiers & & & \\
      TIC ID & TIC 341630071 & TIC 125520907 & 3 \\[0.03cm]
      2MASS ID & J10223416+8305013 & J23402193+3556271 & 2 \\[0.03cm]
      GAIA DR3 ID & 1146601241188708608 & 2879320137933766912 & 1\\[0.03cm] 
      \\
      Astrometric properties & & &\\
        $\alpha$ (J2000)  & 10:22:33.98  & 23:40:21.92  & 1\\[0.03cm] 
        $\delta$ (J2000) & 83:05:01.41 & 35:56:27.02 & 1\\[0.03cm] 
        Parallax (mas) & $3.5201 \pm 0.014$ & $3.7829\pm 0.015$ & 1\\[0.03cm] 
        RUWE & 0.98 & 0.90 & 1 \\[0.03cm]
        \\
        Photometric properties & & & \\
        \textit{TESS} (mag) & $10.705 \pm 0.007$& $10.065 \pm 0.006$ & 3 \\[0.03cm] 
        $B$ (mag) & $11.76 \pm 0.15$& $11.64 \pm 0.16$ & 3 \\[0.03cm]
        $V$ (mag) & $11.10 \pm 0.01$ & $10.791  \pm 0.012$ & 3 \\[0.03cm]
        Gaia$_{BP}$ (mag) & $9.928 \pm 0.003$ & $10.402 \pm 0.003$ & 1 \\[0.03cm]
        Gaia (mag) & $9.612 \pm 0.003$ & $10.142 \pm 0.003$ & 1 \\[0.03cm]
        Gaia$_{RP}$ (mag) & $9.132 \pm 0.004$ & $9.715 \pm 0.004$ & 1 \\[0.03cm]
        $J$ (mag) & $10.131\pm 0.022$ & $9.391 \pm 0.021$ & 2 \\[0.03cm]
        $H$ (mag) & $9.880 \pm 0.029$ & $9.043 \pm 0.021$ & 2 \\[0.03cm]
        $K_S$ (mag) & $9.835 \pm 0.022$ & $8.94 \pm 0.016$ & 2 \\[0.03cm]
        $W_1$ (mag) & $9.764 \pm 0.023$ & $8.884 \pm 0.023$ & 4 \\[0.03cm]
        $W_2$ (mag) & $9.789 \pm 0.02$ & $8.945 \pm 0.02$ & 4 \\[0.03cm]
        $W_3$  (mag) & $9.729 \pm 0.035$ & $8.875 \pm 0.031$ & 4 \\[0.03cm]
        $W_4$  (mag) & $9.11$ & $8.116$ & 4 \\[0.03cm]
        \\
        Derived properties & & & \\
        $T_\mathrm{eff}$ [K]& $5983 \pm 50$ & $5561 \pm 70$ & 5 \\[0.03cm]
        $\log g$ [cgs]& $4.2 \pm 0.1$  & $4.00 \pm 0.1$ & 5 \\[0.03cm] 
        $\left[\textrm{Fe/H}\right]$ [dex] & $-0.29^{+0.07}_{-0.08}$  & $0.04^{+0.07}_{-0.08}$ & 5 \\[0.03cm]
        R$_*$ [R$_\odot$] &  $1.39^{+0.03}_{-0.02}$ & $2.02 \pm 0.03$ & 5 \\[0.03cm] 
        M$_*$ [M$_\odot$] & $0.99 \pm 0.04$  & $1.13 \pm 0.04$ & 5 \\[0.03cm] 
        $v \sin i$ [km/s] & $4.4 \pm 0.5$ & $3.5 \pm 0.5$& 5 \\[0.03cm]
        P$_{rot}/\sin i$ [d] & $16 \pm 2$& $29 \pm 4$ & 5 \\[0.03cm]
       Age [Gyr] & $8.0 \pm 1.0$ & $6.4 \pm 0.6$ & 5 \\[0.03cm] 
       Distance [pc] & $283.4^{+1.6}_{-1.4}$ & $265.2^{+1.9}_{-1.5}$ & 5 \\[0.03cm]
       \hline
    \end{tabular}
    \begin{tablenotes}
       \item References: 1) \textit{Gaia} DR3 \citep{Gaia2020}; 2) 2MASS \citep{2mass2003}; 3) TICv8 \citep{tic2022}; 4) WISE \citep{wise2010}; 5) This work.
       \end{tablenotes}
     \end{threeparttable}
     }
\end{table}

\section{Data analysis} \label{Analysis}

\subsection{TESS photometry}

We first analyzed the photometric data from TESS to identify the transit signals and looked for signs of stellar activity. We also explored whether the photometric data warrant the use of nonzero eccentricity in the modeling via the photo-eccentric effect \citep{dawson2012photoeccentric}. Our analysis of the TESS data using generalized-Lomb-Scargle periodograms \citep[GLS, ][]{Zechmeister2009} shows no significant signs of the stellar rotation periods, indicating the quiet nature of the two stars. For the transit fitting, we used the Python package \texttt{juliet} \citep{espinoza2019juliet}, which employs a nested sampling algorithm via dynesty \citep{speagle2020dynesty}, enabling a comparison of different models through their Bayesian evidence ratios $\Delta \ln(z)$.

Both planets were modeled with a circular orbit and a model where the eccentricity and the argument of periastron ($\omega$) were allowed to vary freely between 0 and 1, and 0$^\circ$ and 360$^\circ$, respectively. We imposed Gaussian priors on the stellar density ($\rho_*$) and the limb-darkening coefficients \citep[parametrized as $q_1$ and $q_2$,][]{Kipping2013} calculated via the ExoCTK package \citep{Bourque2021} from our stellar modeling in Section \ref{stellar}. The rest of the parameters, including the mid-transit time, $T_0$, the orbital period, $P$, the impact parameter, $b$, and the planet-to-star radius ratio, $R_p/R_*$, as well as the offset and noise parameters, were given uniform priors. Comparing the Bayesian evidence of the circular and eccentric transit-only fits, we found a slight preference according to \cite{trotta2008bayes} for the eccentric model for both stars ($\Delta \ln(z) = 2$ for TOI-2147 and $\Delta \ln(z) = 3$ for TOI-6019). 

\subsection{Combined transit and RV fitting}
To derive the final properties of the two planet candidates, we modeled all available photometry together with the RV data from MaHPS using \texttt{juliet}. Again, we fit a circular and an eccentric model to compare their Bayesian evidence. A full list of the fitted parameters and their priors for TOI-2147 and TOI-6019 is given in Tables \ref{tab:toi2147} and \ref{tab:toi6019}.

\begin{figure}
    \centering
    \includegraphics[width=0.8\columnwidth]{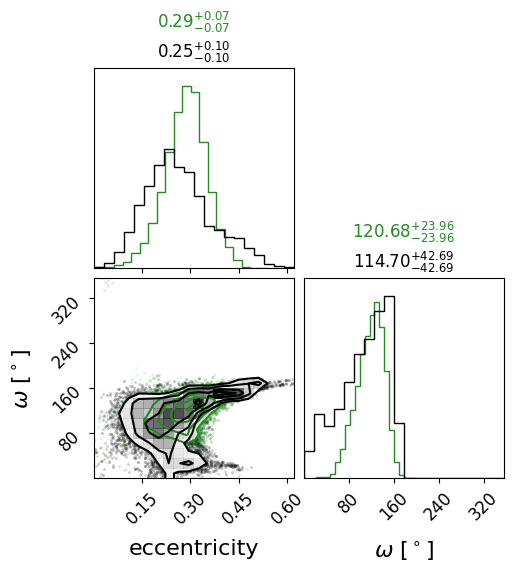} 
    \includegraphics[width=0.8\columnwidth]{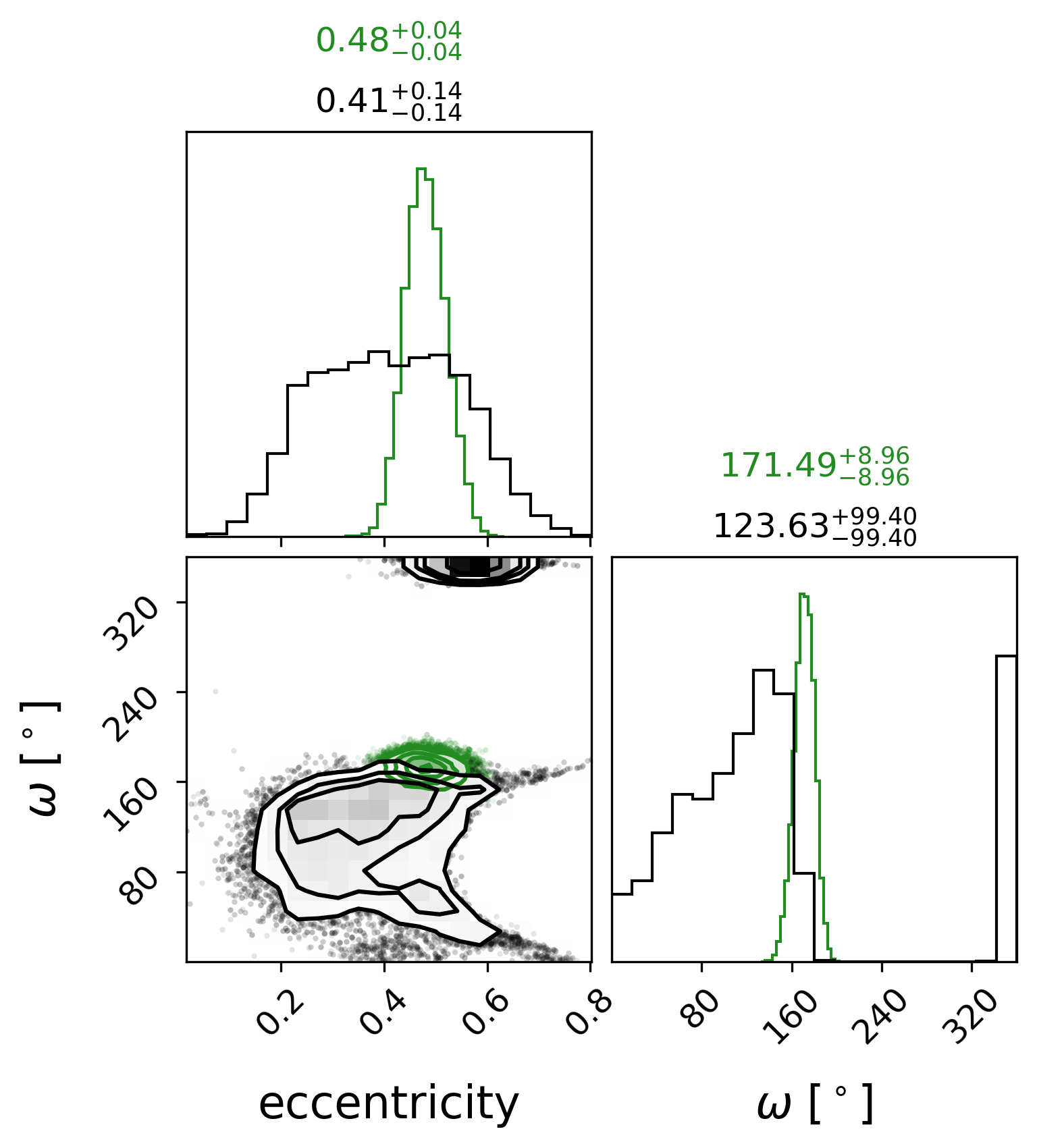} 
    \caption{Posterior distributions for the eccentricity and argument of periastron for TOI-2147 (top) and TOI-6019 (bottom). The black contours are the result of the fitting of the TESS light curves only, while the green contours are derived from the combined fitting of RV and photometry.}
    \label{fig:ecccorner}
\end{figure}

For TOI-2147, the eccentric model is preferred with a $\Delta \ln(z)$ of 5 in the combined RV and transit fits. As shown in Figure \ref{fig:ecccorner}, the solution for $e$ and $\omega$ in our joint fit matches one of the explored areas in the posterior distribution of $e$ and $\omega$ in the transit-only fits. Additionally, our results for the eccentricity satisfy the criterion from \cite{Lucy1971} for the eccentricity to be significant. Therefore, we adopted the values from the eccentric model for the final parameters of TOI-2147\,b.

In the case of TOI-6019, we find very strong evidence for eccentricity in the combined data with a $\Delta \ln(z)$ of 35. While the combination of $\omega$ and $e$ from the combined fit of TOI-6019 is at the edge of the transit-only posteriors, the values agree within their uncertainties (see Figure \ref{fig:ecccorner}). Given the large difference in the Bayesian evidence, we again adopt our results from the eccentric fit. The RV time-series and phase-folded data with the Keplerian model for both planets are shown in Figure \ref{fig:bothrv}. All the light curves are shown in Appendix \ref{alllc}.
\begin{table}
    \centering
    \caption{\label{tab:toi2147}Priors and final parameter values (median and $1\sigma$ confidence intervals) from the \texttt{juliet} fit to the RVs and photometry of TOI-2147.}
    \resizebox{0.99\columnwidth}{!}{
    \begin{threeparttable}[b]
    \centering
    \begin{tabular}{lcc}
    \hline    \hline \\
    Parameter name & Prior & Derived values \vspace{4pt} \\
    \hline  \\
    Fitting parameters \vspace{4pt}  \\
        $P$ [d]        &      $\mathcal{U}[26.0,26.4]$       &  $26.20518 \pm 0.00003$  \vspace{2pt} \\
        $T_0$ (BJD$_{\rm TDB} -2457000$)   & $\mathcal{U}[2389.45 ,2389.6]$  & $2389.5349  ^{+0.0006}_{-0.0007}$ \vspace{2pt} \\
        $R_p / R_*$ & $\mathcal{U}[0.001, 1]$ & $0.0694^{+0.0009}_{-0.0008}$ \vspace{2pt}\\
        $b$ & $\mathcal{U}[0, 1]$& $0.33^{+0.14}_{-0.17}$ \vspace{2pt}\\
        $e$ & $\mathcal{U}[0, 0.99]$ & $0.29 \pm 0.07$ \vspace{2pt}\\
        $\omega$ [$^\circ$] & $\mathcal{U}[0, 360]$& $121 \pm 30$ \vspace{2pt} \\
        $K$ [m~s$^{-1}$]      & $\mathcal{U}[0,200]$       &  $26 \pm 5$ \vspace{2pt} \\
        $\rho_*$ [kg/$\text{m}^3$]& $\mathcal{N}[520, 40.]$ & $520 \pm 40$ \vspace{6pt}\\
        Instrumental parameters \vspace{4pt}  \\
        $q_{1, TESS}$ & $\mathcal{N}[0.28,0.1]$ & $0.27^{+0.08}_{-0.07}$ \vspace{2pt}\\
        $q_{2, TESS}$ & $\mathcal{N}[0.26,0.1]$ & $0.24 \pm 0.09$ \vspace{2pt}\\
        $\mu_{\rm MaHPS}$ (m s$^{-1}$)  & $\mathcal{U}[-100,100]$   & $5 \pm 4$ \vspace{2pt}\\
        $\sigma_{\rm MaHPS}$ (m s$^{-1}$)  & $\mathcal{L}[0.001,100]$   & $25.7^{+3.0}_{-2.9}$\vspace{4pt}\\
        \hline
        \\
        TOI-2147\,b properties \vspace{4pt} \\
        $a/R_*$ & & $26.6^{+0.6}_{-0.7}$\vspace{2pt} \\
        $a$ [au] & & $0.172 \pm 0.006$ \vspace{2pt}\\
        $T_{14}$ [hours] & & $5.85\pm 0.12$ \vspace{2pt}\\
        $i$ [$^\circ$] & & $89.3 \pm 0.4$\vspace{2pt}\\
        $R_p$ [$\mathrm{R}_\oplus$] & & $10.5 \pm 0.3$ \vspace{2pt}\\
        $M_p$ [$\mathrm{M}_\oplus$] & & $116 \pm 22$\vspace{2pt}\\
        $\rho_p$ [kg/m$^3$] & & $550 \pm 110$\vspace{2pt}\\
        $T_{eq}$ [K] & & $820 \pm 20$ \vspace{4pt}\\
    \hline 
    \end{tabular}
    \begin{tablenotes}
       \item Notes: $\mathcal{U}$ indicates a uniform prior, while $\mathcal{N}$ and $\mathcal{L}$ indicate normal and log-normal priors, respectively.
       \end{tablenotes}
     \end{threeparttable}
     }
\end{table}

\begin{table}
    \centering
    \caption{\label{tab:toi6019}Priors and final parameter values (median and $1\sigma$ confidence intervals) from the \texttt{juliet} fit to the RVs and photometry of TOI-6019.}
    \resizebox{0.99\columnwidth}{!}{
    \begin{threeparttable}[b]
    \centering
    \begin{tabular}{lcc}
    \hline \hline   \\
    Parameter name & Prior & Derived values  \vspace{4pt}    \\
    \hline\\
    Fitting parameters \vspace{4pt}    \\
        $P$ [d]        &      $\mathcal{U}[14.4,14.6]$       &  $14.547952^{+0.000036}_{-0.000035}$  \vspace{2pt} \\
        $T_{0}$ (BJD$_{\rm TDB} -2457000$)   & $\mathcal{U}[2862.1,2862.5]$  & $2862.3434 \pm 0.0015$  \vspace{2pt}  \\
        $R_{p} / R_*$ & $\mathcal{U}[0.001, 1]$ & $0.0560 \pm 0.0011$  \vspace{2pt} \\
        $b$ & $\mathcal{U}[0, 1]$& $0.72^{+0.05}_{-0.07}$ \vspace{2pt} \\
        $e$ & $\mathcal{U}[0, 0.99]$ &  $0.48 \pm 0.04$ \vspace{2pt} \\
        $\omega$ [$^\circ$]& $\mathcal{U}[0, 360]$&  $171^{+9}_{-8}$ \vspace{2pt}\\
        $K$ [m~s$^{-1}$]      & $\mathcal{U}[0,200]$       &  $41 \pm 4$  \vspace{2pt} \\
        $\rho_*$ [kg/$\text{m}^3$]& $\mathcal{N}[190, 10.]$ & $191 \pm 10$\vspace{6pt} \\
        Instrumental parameters \vspace{4pt} \\
        $q_{1, \rm TESS}$ & $\mathcal{N}[0.34,0.1]$ & $0.32^{+0.09}_{-0.08}$ \vspace{2pt}\\
        $q_{2, \rm TESS}$ & $\mathcal{N}[0.32,0.1]$ & $0.3 \pm 0.1$ \vspace{2pt}\\
        $\mu_{\rm MaHPS}$ (m s$^{-1}$)  & $\mathcal{U}[-100,100]$   & $8.6^{+2.8}_{-2.6}$ \vspace{2pt}\\
        $\sigma_{\rm MaHPS}$ (m s$^{-1}$)  & $\mathcal{L}[0.001,100]$   & $17.6^{+2.4}_{-2.2}$ \vspace{4pt} \\
        \hline \\
        TOI-6019\,b properties \vspace{4pt}  \\
        $a/R_*$ & & $12.9 \pm 0.3$ \vspace{2pt}\\
        $a$ [au] & & $0.1212 \pm 0.0033$ \vspace{2pt}\\
        $T_{14}$ [hours] & & $5.5 \pm 0.1$ \vspace{2pt}\\
        $i$ [$^\circ$] & & $86.80 \pm 0.32$\vspace{2pt}\\
        $R_p$ [$\mathrm{R}_\oplus$] & & $12.3 \pm 0.3$\vspace{2pt} \\
        $M_p$ [$\mathrm{M}_\oplus$] & & $149 \pm 15$\vspace{2pt}\\
        $\rho_p$ [kg/m$^3$] & & $440 \pm 60$\vspace{2pt}\\
        $T_{eq}$ [K] & & $1095 \pm 30$\vspace{4pt} \\
        
    \hline 
    \end{tabular}
        \begin{tablenotes}
       \item Notes: $\mathcal{U}$ indicates a uniform prior while $\mathcal{N}$ and $\mathcal{L}$ indicate normal and log-normal priors respectively.
       \end{tablenotes}
       \end{threeparttable}
    }
\end{table}

\begin{figure*}
    \centering
    \includegraphics[width=0.99\columnwidth]{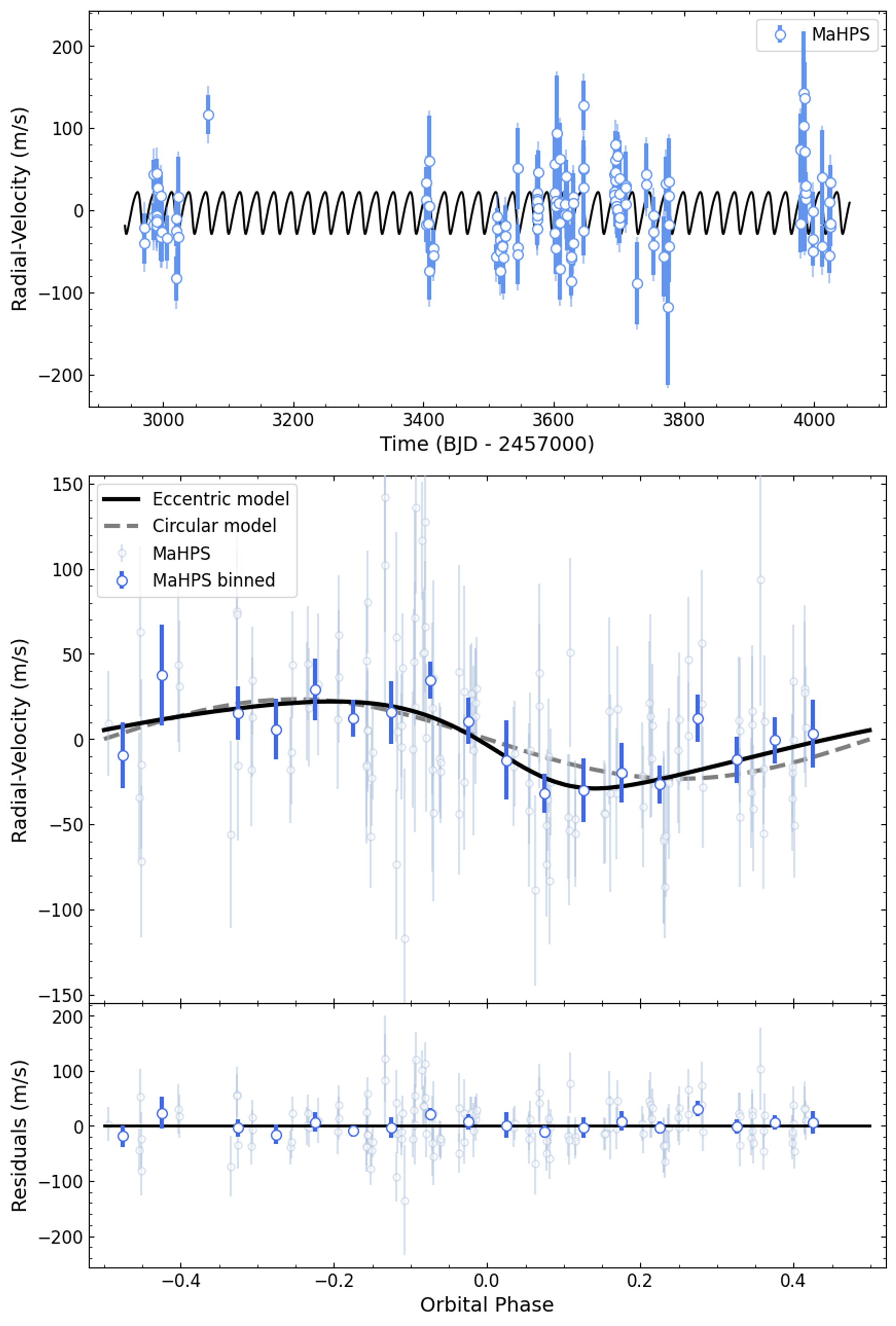} 
    \includegraphics[width=0.99\columnwidth]{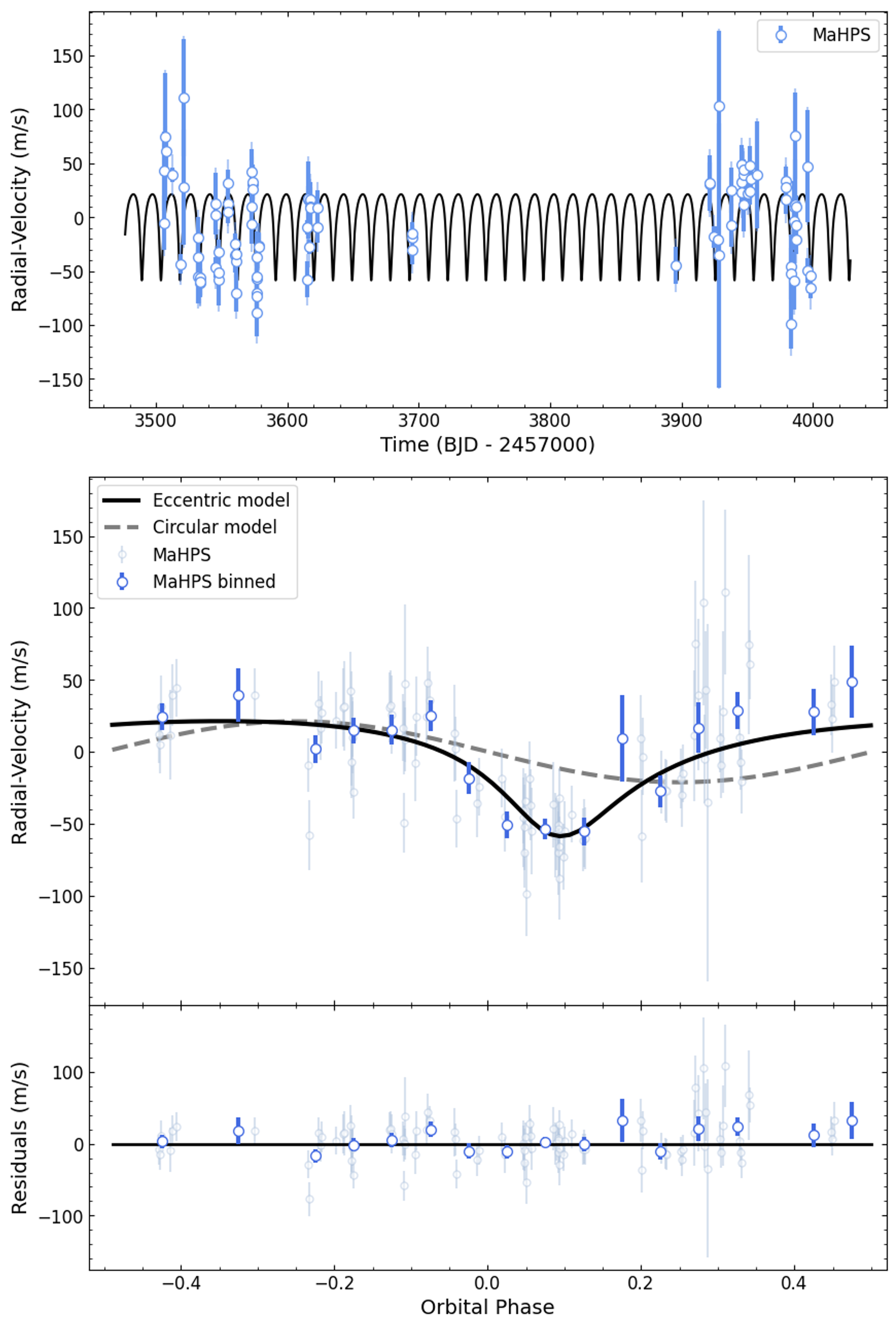} 
    \caption{Results of the \texttt{juliet} fit for TOI-2147 (left) and TOI-6019 (right). First panel: Time-series radial velocity curve with the best-fit Keplerian model in black. Second panel: RVs phase-folded to the period of the planets. The dark blue data points represent the MaHPS RV data binned at 0.05 phases. Third Panel: Residuals after subtracting the eccentric model of the planet.}
    \label{fig:bothrv}
\end{figure*}

\section{Results and discussion} \label{results_disc}

\subsection{Planetary parameters}
\begin{figure}
    \centering
    \includegraphics[width=0.99\columnwidth]{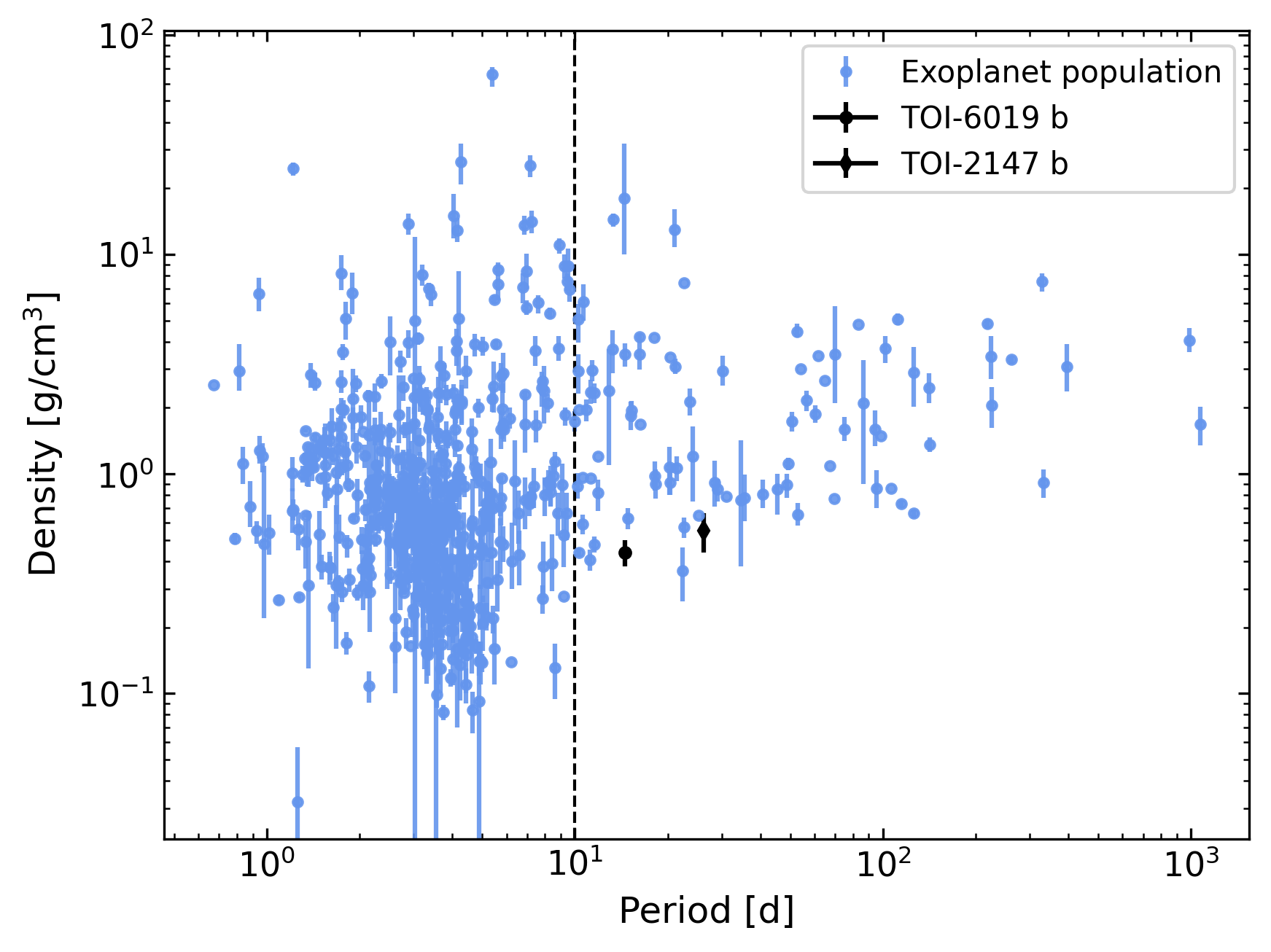} 
    \caption{Density as a function of period for giant planets ($M_p > 0.3~M_J$) with mass and radius measured to a precision of $33\%$ or better, taken from the NASA Exoplanet Archive \citep{Christiansen2025}. TOI-6019\,b and TOI-2147\,b are highlighted in black. The vertical dotted line indicates the separation between the hot and warm Jupiter populations.}
    \label{fig:dens}
\end{figure}

We have confirmed the planetary nature of the two WJs TOI-2147\,b and TOI-6019\,b. TOI-2147\,b is a planet with a radius of $10.5 \pm 0.3~\mathrm{R}_\oplus$, a mass of $116 \pm 22~\mathrm{M}_\oplus$, orbiting its host star every 26.2 days. TOI-6019\,b has a radius of $12.3 \pm 0.3~\mathrm{R}_\oplus$, a mass of $149 \pm 15~\mathrm{R}_\oplus$, and has a slightly evolved sub-giant host star which it orbits every 14.5 days. The radius of TOI-6019\,b is among the largest of the warm ($P>10$~days) planet population. Both planets have lower bulk densities than Jupiter ($550 \pm110$\,kg/m$^3$ and $440 \pm60$\,kg/m$^3$), indicating that they are slightly inflated, although their densities are not as low as some of the very puffy HJs \citep[e.g., WASP-107\,b;][]{Anderson2017}. The densities of the known giant planet population are shown in Figure \ref{fig:dens}. While our two planets are too far away from their host stars and too mature to have significant thermal inflation, they could be subject to tidal heating and inflation given their moderate eccentricities. 

\subsection{Formation and migration history}
\begin{figure}
    \centering
    \includegraphics[width=0.99\columnwidth]{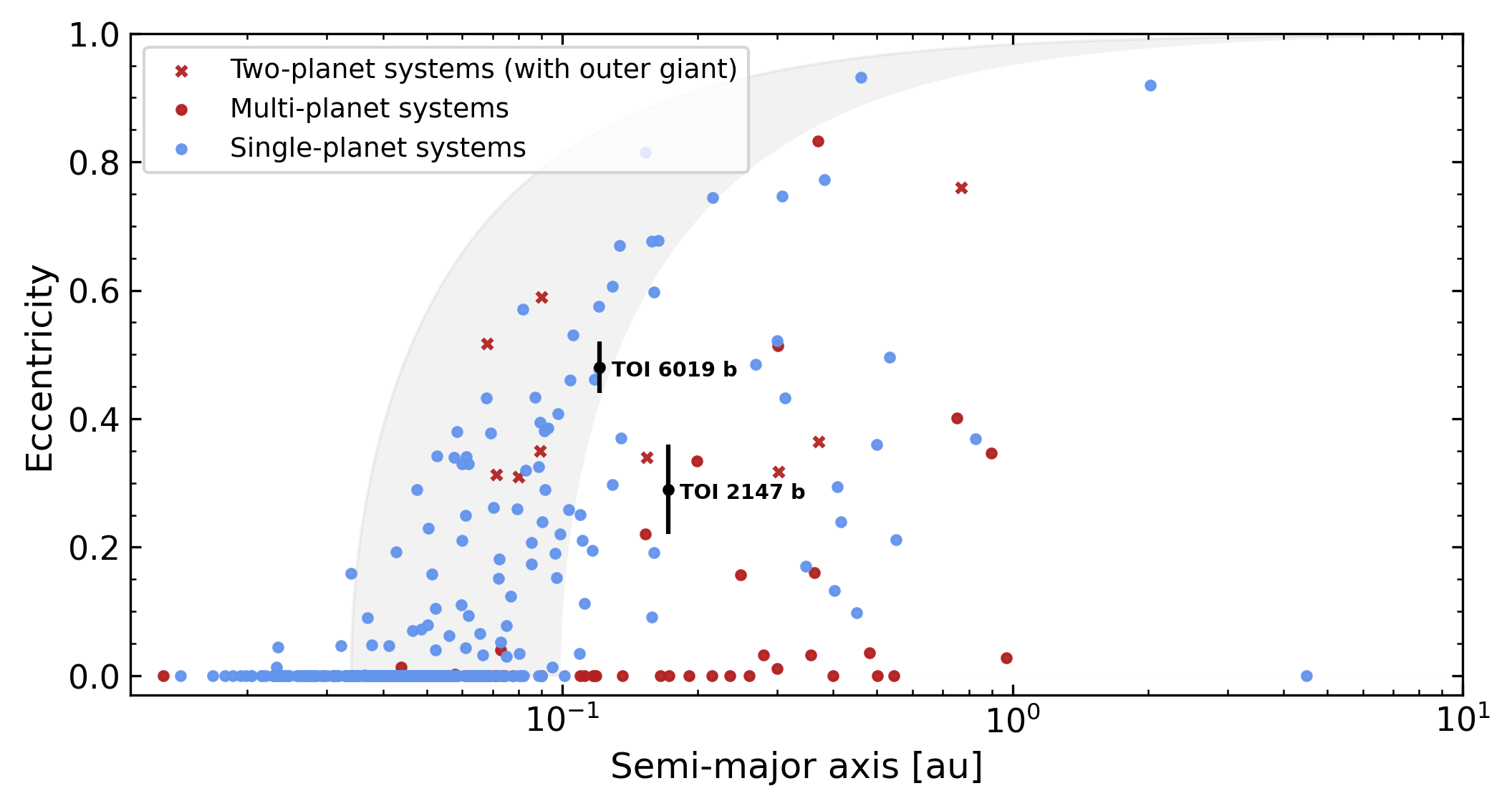} 
    \caption{Population of giant planets ($M_p > 0.3~M_J$) with true mass measurements and measured eccentricities ($\sigma_e/e<0.33$) taken from NASA Exoplanet Archive \citep{Christiansen2025}. Planets where the eccentricity was set as 0 without an associated uncertainty are also included. The gray region shows constant angular momentum tracks of planets expected to end up as HJs \citep[adopted from ][]{Dawson2018}. Planets without known companions are colored in blue markers, while giant planets in multiplanet systems are plotted with red markers. Giant planets with exactly one detected outer giant companion ($M_p>0.3~M_J$ and $P_{orb}>1000~$d for the companion) are indicated by the red cross. TOI-2147\,b and TOI-6019\,b are highlighted in black.}
    \label{fig:ecc}
\end{figure}

The eccentricities of TOI-2147\,b ($e = 0.29 \pm 0.07$) and TOI-6019\,b ($e=0.48^{+0.05}_{-0.04}$) place them among the more dynamically excited WJs, potentially offering insights into their formation and subsequent evolution. Overall, WJs exhibit a wider range of orbital eccentricities than their hot Jupiter counterparts \citep{Dawson2018,Morgan2026}, where the majority are found on low eccentricity orbits. Figure \ref{fig:ecc} shows all known giant planets ($M_p > 0.3~M_J$) with true mass measurements and measured eccentricities ($\sigma_e<0.2$). The gray region shows constant angular momentum tracks which follow $a_{final}=a(1-e^2)$ as expected for planets undergoing high-eccentricity migration. Planets inside this region are considered proto-HJs as their orbits will shrink over time, bringing the planet close to its host star ($a<0.1$~AU). TOI-6019\,b lies just inside this gray region and is predicted to have a final semi-major axis of $0.093\pm0.007$~AU after circularizing (i.e., a period of $10 \pm 1$~days). 

The broad eccentricity distribution of WJs is indicative of multiple origin channels. Low-eccentricity WJs likely migrated via disk migration or might have even formed in situ. On the other hand, the high-eccentricity WJs have likely undergone a more dynamical migration mechanism, such as planet-planet scattering or Zeipel-Lidov-Kozai (ZLK) oscillation. Another indicator for the dynamical history of a giant planet is the existence of nearby companions. Giant planets in single-planet systems are colored in blue in Figure \ref{fig:ecc}, while multiplanet giants are plotted with red markers. The giant planets that only have one, massive outer companion ($M_p>0.3~M_J$ and $P_{orb}>1000~$d for the companion) are indicated by the red crosses. The majority of the HJs are found in single-planet systems and the ones that are in multiplanet systems tend to have a long-period giant companion. This is consistent with the high-eccentricity formation channel, where the outer giant scatters the hot Jupiter, clearing any close companion planets during the migration.

Again, WJs tend to show more diverse system architectures, with many of the low to medium eccentricity planets ($e<0.6$) residing in multiplanet systems, indicating a more dynamically quiet evolution. To further investigate possible formation channels of our two planets, we searched for signals of additional planets in the spectroscopic and photometric data. A periodogram analysis of the RV datasets of TOI-2147\,b and TOI-6019\,b finds no significant periodicities apart from the known planets. To quantify the significance of the absence of such signals, we performed an injection recovery test to check the sensitivity of our RV data. Using the \texttt{RVSearch} python package \citep{Rosenthal2021} to inject 3000 synthetic planets drawn from a log-uniform period (1-1000~days) and RV amplitude (1-500~m/s) distribution into our RV data. We then used the $\Delta$BIC periodogram search implemented in \texttt{RVSearch} to see if we could recover these planets. The resulting completeness maps are shown in Appendix \ref{kap:compl}. Based on these completeness maps, we can only rule out the existence of giant planet companions ($M_p \geq 100~\mathrm{M}_\oplus$) with a period up to 100~days. Additionally, we looked for transit timing variations in the \textit{TESS} light curves of both planets. For this, we fit the transits separately using \texttt{Juliet} to get a transit mid time for each individual transit and then subtracted the expected mid transit time from our combined fit. The resulting O-C diagrams are shown in Figure~\ref{fig:oc}. The measured transit times show some small scatter with an amplitude of $\sim 5$~minutes. To look for periodic signals suggesting the presence of undetected companions, we performed a periodogram analysis with a Lomb-Scargle periodogram, which revealed no periodicities, indicating that the scatter might be coming from sources other than a planet (e.g., systematics, stellar activity). Although this has been attributed to the small number of transits, no definitive inferences can be made. The elevated eccentricities and absence of companion planets are consistent with our two planets undergoing high-eccentricity migration; however, due to the limited sensitivity of our RV (and TTV) data, we cannot rule out the existence of smaller companions, which would imply a more quiescent migration.
\\

\subsection{Interior structure modeling}
From the derived properties of TOI-2147\,b and TOI-6019\,b, we model their interior structure by comparing them to an extended grid of interior structure models from \cite{Thomas20252} that was calculated with the GAS gianT modeL for Interiors \citep[GASTLI][]{acuna2021,acuna2024}. GASTLI was developed to model the interior structure of gas-giant and sub-Neptune exoplanets by assuming a core composed of water and rock with a fully convective envelope composed of H/He and metals on top. The interior is coupled self-consistently to a grid of atmospheric models. \cite{Thomas20252} used GASTLI's default atmospheric grid for warm exoplanets with equilibrium temperatures of $T_{\rm eq} \leq 1000$\,K.

For both planets, we ran retrievals on the grid with Gaussian priors on the derived radius, mass, and ages. Following the procedure in \cite{Thomas20252}, we ran three retrievals for both planets, assuming different levels of envelope enrichment as we have no information on the atmospheric metallicity ($\log(\mathrm{Fe}/\mathrm{H})$) of our planets. The three different metallicity cases are: low ($ -2.0 < \log(\mathrm{Fe}/\mathrm{H}) < 0.5$), medium ($ 0.5 < \log(\mathrm{Fe}/\mathrm{H}) < 1.4$), and high ($ 1.4 < \log(\mathrm{Fe}/\mathrm{H}) < 1.7$). Interestingly, we found that for our high atmospheric metallicity retrieval ($1.4<\log(\mathrm{Fe}/\mathrm{H})<1.7$), the model was not able to match the observed radius with the observed mass. This indicates that, based on the observed mass and radius of the two planets, we would not expect any significant metal enrichment of their atmospheres, which is expected based on observations of the solar system giants and other gas giant exoplanets \citep{kipping2014characterizing,Thorngren2016,Ohno2025,Thomas20252}.
Table \ref{tab:env} shows our derived bulk metal mass fraction ($Z_p$) and envelope mass fraction ($f_{env}$) for the other two retrievals, which were largely consistent with each other. 

Given the moderate eccentricities and low bulk densities, there is a possibility that the radii of our two planets are inflated, which can lead to an overestimation of the envelope mass fraction in interior modeling \citep{Millholland2020,Hallatt2026}. Tidal inflation is stronger for planets close to their host star, but could still play a role for eccentric WJs. We can use Eqs. 23 and 24 in \cite{Leconte2010} to estimate the amount of tidal heating that is expected from the derived eccentricity \citep[see also][]{acuna2025}. The reduced tidal quality factor of $Q^\prime_p=3Q_p/2k_2$ is $\sim 10^5$ \citep{Yoder1981,Lainey2009} for Jupiter, $\sim 10^4 -10^5$ for Uranus and Neptune \citep{Tittemore1990,Banfield1992}, and $\sim 10^3-10^4$ for Saturn \citep{Lainey2012,Polycarpe2018}. Constraints on $Q^\prime_p$ for HJs have generally resulted in larger values $Q^\prime_p \gtrsim10^5$ \citep[see e.g., ][]{Quinn2014,Kawai2025}. As we had no information on $Q^\prime_p$ for our two planets, we took the range $Q^\prime_p = 10^3-10^6$ to estimate the minimum and maximum expected temperature from tidal heating. For TOI-2147\,b this results in a tidal temperature of $T_{tide} = 40-230~$K and for TOI-6019\,b, we estimated $T_{tide} = 140-800~$K. With this, we run another set of interior retrievals where we put a prior on the internal temperature from the tidal heating. The total internal temperature is calculated as $T_{int}=(T_{tide}^4 + T_{age}^4)^{1/4}$ with $T_{age}$ taken from the previous internal structure retrievals ($\sim 60$~K for TOI2147\,b and $\sim70$~K for TOI-6019\,b). This results in a prior on the internal temperature of TOI-2147\,b between 65~K and 230~K and between 145~K and 800~K for TOI-6019\,b. The results of this retrieval are summarized in Table \ref{tab:env}. Comparing the results of this interior retrieval with the previous run highlights the importance of including tidal heating even for longer-period planets. Considering only the internal heat from secular cooling, both planets are inferred to have large envelope mass fractions comparable to Jupiter ($f_{env} > 0.9$). However, when taking into account the additional heat from tidal effects, the inferred $f_{env}$ drops significantly and is more comparable to Saturn's envelope mass fraction, which is in line with the measured masses of the two planets being closer to Saturn's than Jupiter's. Measurements of the atmospheric composition, particularly the bulk atmospheric metallicity, would allow us to improve the accuracy of the interior modeling and give a more detailed picture of the nature of these two planets.

\begin{table}
    \centering
    \caption{\label{tab:env}Results of the interior retrievals for TOI-2147\,b and TOI-6019\,b.}
    \resizebox{\columnwidth}{!}{
    \begin{tabular}{lcccc}
    \hline
    \hline 
    \multicolumn{1}{c}{} & \multicolumn{2}{c}{TOI-2147} & \multicolumn{2}{c}{TOI-6019} \\
         & $f_{env}$ & $Z_p$ & $f_{env}$ &  $Z_p$\\
      \hline
      \\
      \textit{No Tidal Heating}\\
      \\
      low $\log(\mathrm{Fe}/\mathrm{H})$& $0.93^{+0.04}_{-0.05}$ & $0.08^{+0.05}_{-0.04}$ & $0.99^{+0.01}_{-0.02}$& $0.02^{+0.02}_{-0.01}$ \vspace{0.05cm}\\
      medium $\log(\mathrm{Fe}/\mathrm{H})$& $0.97^{+0.02}_{-0.04}$&$0.12^{+0.04}_{-0.03}$ &$0.99^{+0.01}_{-0.01}$& $0.08^{+0.02}_{-0.01}$ \vspace{0.05cm}\\
      \\
      \textit{With tidal heating}\\
      \\
      low $\log(\mathrm{Fe}/\mathrm{H})$& $0.75\pm 0.05$& $0.26\pm 0.05$ &$0.65^{+0.06}_{-0.11}$ & $0.35^{+0.11}_{-0.05}$ \vspace{0.05cm} \\
      medium $\log(\mathrm{Fe}/\mathrm{H})$&$0.77^{+0.12}_{-0.07}$ & $0.32\pm 0.05$ & $0.76^{+0.12}_{-0.17}$ & $0.33^{+0.15}_{-0.10}$ \vspace{0.05cm}\\
      high $\log(\mathrm{Fe}/\mathrm{H})$&$0.98^{+0.01}_{-0.02}$ & $0.30\pm 0.01$ & $0.88^{+0.08}_{-0.10}$& $0.43^{+0.05}_{-0.06}$ \vspace{0.05cm} \\
    \hline
    \end{tabular}
    }
\end{table}

\subsection{Prospects for atmospheric follow-up}

Studying the C/O ratio of gas giant atmospheres may provide important constraints on their formation pathways, as it traces the relative contributions of solid and gas accretion at different locations in the protoplanetary disc \citep[e.g.][]{Oeberg_2011,Madhusudhan_2011}. The relatively low bulk densities of TOI-2147\,b and TOI-6019\,b make both planets promising targets for atmospheric follow-up to probe their chemical composition. However, their transmission spectroscopy metrics are 44 and 40, respectively, which makes them not ideal targets for follow-up using transmission spectroscopy where the threshold for favorable targets is assumed to be 90 \citep{Kempton2018}. However, this threshold does depend on the telescope aperture and instrument capabilities. Future facilities such as the ELT are expected to be able to probe targets with substantially lower TSM values \citep{Palle2025}.

For the ground-based follow-up, the feasibility for characterization strongly depends on the sky position and orbital configuration. Both planets are on eccentric orbits with favorable transit geometry ($\omega < 180\deg$) for the transmission spectroscopy \citep{Prinoth_2024}. However, the TOI-2147 system is circumpolar with a declination of +83:05:01.41 (see Table~\ref{tab:star}) and consequently never reaches sufficiently low airmass ($\leq 2.8$) for extended periods from major Northern Hemisphere observatories equipped with $\geq 8$m mirrors. This limits its accessibility for high-precision ground-based time-series observations.

In contrast, TOI-6019\,b is well suited for atmospheric follow-up from the ground and is overall the more favourable system given its shorter orbital period and brighter host star. With an equilibrium temperature of $\approx1100~$K, its atmosphere is expected to be H/He-dominated with prominent molecular absorbers such as \ch{H2O}, \ch{CH4}, and \ch{CO}, making it an excellent target for near-infrared spectroscopy.

Therefore, we explored the detectability of its atmosphere using Gemini-N/IGRINS2 \citep{Lee_2022,Oh_2024}, which simultaneously covers the H and K bands at a spectral resolving power of $\mathcal{R} \sim 45,000$. Based on the stellar parameters in Table~\ref{tab:star}, we adopted exposure times of 300~s with a readout time of 39.5~s, yielding a signal-to-noise ratio (S/N) of $\approx150$ per nodding position, as predicted using the integration time calculator for IGRINS2\footnote{\href{https://www.gemini.edu/instrumentation/igrins-2/exposure-time-estimation\#ITC}{https://www.gemini.edu/instrumentation/igrins-2/exposure-time-estimation
\#ITC}}.

We simulated transit observations using \texttt{ExoAtmoSim} \citep{bibiana_prinoth_2024_11505486, Prinoth_2024}, which wraps \texttt{tayph} \citep{hoeijmakers_2024_11506199}, \texttt{petitRADTRANS} \citep{Molliere_2019}, and \texttt{FastChem Cond} \citep{Kitzmann_2023} into an end-to-end simulator to generate time-series observations of exoplanets for a variety of instruments. We assumed the planetary atmospheres to be in chemical and hydrostatic equilibrium, with an isothermal temperature-pressure profile at the equilibrium temperatures listed in Table~\ref{tab:toi6019} and metallicity of the host stars listed in Table~\ref{tab:star}. The pressure at the bottom of the atmosphere was set to 10~bar. We included line-by-line opacities from \ch{H2O} \citep{Polyansky_2018}, \ch{CO} \citep{Li_2015} and \ch{CH4} \citep{Yurchenko_2024}, and continuum-opacities through collisionally-induced absorption of \ch{H2}-\ch{H2} and \ch{H2}-\ch{He} \citep{Richard_2012}, as well as Rayleigh scattering of \ch{H2} and \ch{He}. The atmosphere was assumed to be cloud-free. We modeled the stellar spectra with a PHOENIX spectrum \citep{husser2013} adopting the stellar parameters in Table~\ref{tab:star}, including telluric contamination using {\tt telfit} \citep{telfit}, and assuming a precipitable water vapor of 2.0~mm, which is slightly worse than average for Mauna Kea \citep{Otarola_2015}. For an extensive description of the simulator and radial velocities of the components, see \citet{Prinoth_2024}.

We then performed cross-correlation analysis on the simulated transits with \texttt{POSEIDON} \citep{MacDonald_2017,MacDonald_2023,Wang_2025}, applying three, five, and ten principal components using its SYSREM implementation. As \texttt{POSEIDON} assumes circular orbits by default, we modified the code to handle eccentric orbits using \texttt{radvel} \citep{fulton2018radvel} and \texttt{PyAstronomy's} modelsuite \citep{pya} to compute expected radial velocities.

Owing to its relatively long transit duration of 5.2~h (Table~\ref{tab:toi6019}), this system approaches the practical limit of what can be observed during a single night. We therefore consider a more flexible strategy based on partial transit observations. Observing half the transit together with pre-transit baseline provides sufficient temporal coverage to remove stellar and telluric contamination \citep{Cheverall_2024}.

\begin{figure}
    \centering
    \includegraphics[width=\linewidth]{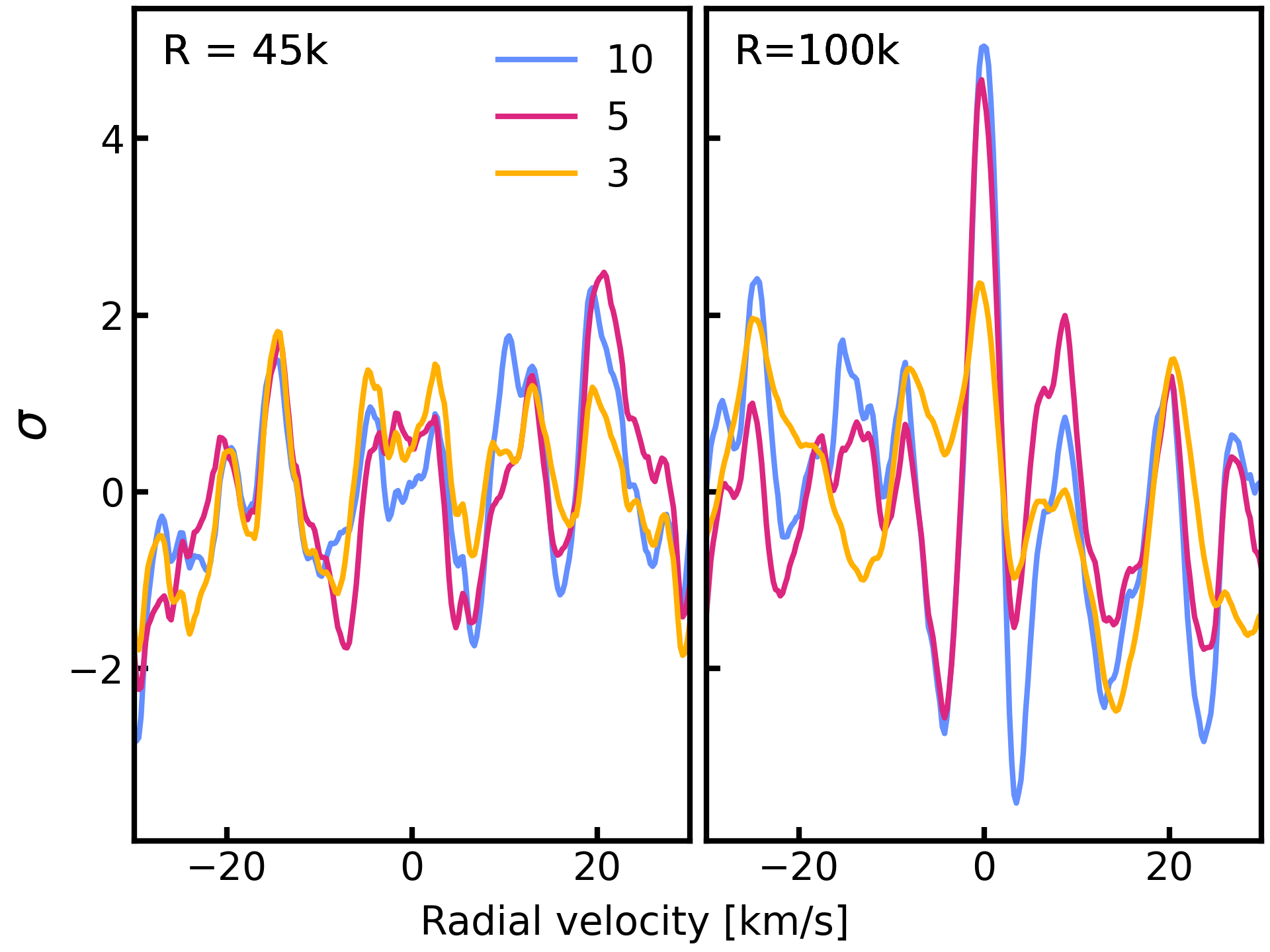}
    \caption{Predicted cross-correlation signal after 2 transits for IGRINS2 (left) and a made-up instrument covering the same wavelength region, but at a resolving power of $\mathcal{R} \sim 100,000$. The number of SYSREM components is shown in different colours. Three components are not sufficient to effectively remove stellar and telluric residuals, but with five or ten, the signal remains the same in the $\mathcal{R} \sim 100,000$-case. No detection has been predicted for $\mathcal{R} \sim 45,000$ (IGRINS2).}
    \label{fig:atmospheric_det}
\end{figure}

At the native resolving power of IGRINS2, telluric contamination remains the dominant limitation. Even when combining five transits (ten simulated observations), the atmospheric signal remains far below the $3\sigma$-level (Figure~\ref{fig:atmospheric_det}). However, when artifically increasing the resolving power to $\mathcal{R} \sim 100,000$, we predict that this would give us the ability to detect the atmosphere at $\gtrsim4\sigma$ with only two transits (via four simulated observations). This demonstrates the need for higher resolving powers when probing these kinds of planetary atmospheres. It also shows that detectability is not necessarily limited by the sensitivity and, instead, it could be affected by the spectral resolution and how successfully telluric contamination can be removed.

\section{Summary} \label{sum}

We present the detection and characterization of two new eccentric WJs, TOI-2147\,b and TOI-6019\,b, using TESS photometry and high-resolution spectroscopy from MaHPS.

\begin{enumerate}

    \item We confirm the planetary nature of TOI-2147\,b and TOI-6019\,b via a joint modeling of the TESS light curves and MaHPS radial velocities. TOI-2147\,b has a radius of $10.5 \pm 0.3\,\mathrm{R}_\oplus$ and a mass of $116 \pm 22\,\mathrm{M}_\oplus$, orbiting its host star every 26.2 days. TOI-6019\,b has a radius of $12.3 \pm 0.3\,\mathrm{R}_\oplus$ and a mass of $149 \pm 15\,\mathrm{M}_\oplus$, orbiting a mildly evolved sub-giant host star with a period of 14.5 days. Both planets have bulk densities slightly below that of Jupiter, placing them among the population of mildly inflated WJs.

    \item Both planets have significantly nonzero eccentricities, $e = 0.29 \pm 0.07$ for TOI-2147\,b and $e = 0.48^{+0.05}_{-0.04}$ for TOI-6019\,b, as evidenced by strong Bayesian evidence in favour of eccentric solutions in the combined photometric and RV fits. TOI-6019\,b is a proto-hot Jupiter whose orbit is expected to circularize to a final semi-major axis of $0.093 \pm 0.007\,\mathrm{au}$.

    \item No additional planetary companions have been detected in the RV periodograms or from transit timing variations. The elevated eccentricities combined with the absence of nearby companions are consistent with a high-eccentricity migration origin for both planets, although the limited sensitivity of our RV data cannot rule out smaller companions that would be consistent with a more quiescent migration scenario.

    \item Interior structure modeling using \texttt{GASTLI} shows that the observed masses and radii of both planets are inconsistent with high atmospheric metallicity, when only considering internal heat based on the ages of the systems. However, when tidal heating is taken into account, the inferred envelope mass fractions are reduced significantly and even higher atmospheric metallicities produce results that are consistent with the observed masses and radii.

\end{enumerate}

\section{Data availability} \label{Sec:DataAvailability}
Tables 5 and 6 are only available in electronic form at the CDS via anonymous ftp to cdsarc.u-strasbg.fr (130.79.128.5) or via http://cdsweb.u-strasbg.fr/cgi-bin/qcat?J/A+A/. The tables contain the following information: column 1 lists the time of the observations, column 2 gives the derived RV values, column 3 gives the error of the RV, and column 4 lists the name of the instrument.

\begin{acknowledgements}

LT acknowledges support from the Excellence Cluster ORIGINS funded by the Deutsche Forschungsgemeinschaft (DFG, German Research Foundation) under Germany's Excellence Strategy – EXC 2094/2 – 390783311.
\\

The Wendelstein 2.1 m telescope project was funded by the Bavarian government and by the German Federal government through a common funding process. Part of the 2.1 m instrumentation, including some of the upgrades for the infrastructure, were funded by the Cluster of Excellence “Origin of the Universe” of the German Science foundation DFG.
\\

Funding for the \textit{TESS} mission is provided by NASA's Science Mission Directorate. KAC and CNW acknowledge support from the \textit{TESS} mission via subaward s3449 from MIT.
\\
We acknowledge the use of public \textit{TESS} data from pipelines at the TESS Science Office and at the TESS Science Processing Operations Center. 
\\
Resources supporting this work were provided by the NASA High-End Computing (HEC) Program through the NASA Advanced Supercomputing (NAS) Division at Ames Research Center for the production of the SPOC data products.
\\
This paper includes data collected by the \textit{TESS} mission, which are publicly available from the Mikulski Archive for Space Telescopes (MAST) operated by the Space Telescope Science Institute (STScI).
\\
This research has made use of the Exoplanet Follow-up Observation Program (ExoFOP; DOI: 10.26134/ExoFOP5) website, which is operated by the California Institute of Technology, under contract with the National Aeronautics and Space Administration under the Exoplanet Exploration Program.
\\
This work makes use of observations from the LCOGT network. Part of the LCOGT telescope time was granted by NOIRLab through the Mid-Scale Innovations Program (MSIP). MSIP is funded by NSF.
\\

This research has made use of the NASA Exoplanet Archive, which is operated by the California Institute of Technology, under contract with the National Aeronautics and Space Administration under the Exoplanet Exploration Program.
\\

This article is based on observations made with the MuSCAT2
instrument, developed by ABC, at Telescopio Carlos Sánchez operated on the island of Tenerife by the IAC in the Spanish Observatorio del Teide.
\\
This work is partly supported by JSPS KAKENHI Grant Numbers JP24H00017, JP24K00689 and JSPS Bilateral Program Number JPJSBP120249910.
\\
We acknowledge financial support from the Agencia Estatal de Investigaci\'on of the Ministerio de Ciencia e Innovaci\'on MCIN/AEI/10.13039/501100011033 and the ERDF “A way of making Europe” through projects PID2021-125627OB-C32 and PID2024-158486OB-C32.
\\

JVH is funded by the Deutsche Forschungsgemeinschaft (DFG, German Research Foundation) - 562755121.
\\

F. M. acknowledges the financial support from the Agencia Estatal de Investigaci\'{o}n del Ministerio de Ciencia, Innovaci\'{o}n y Universidades (MCIU/AEI) through grant PID2023-152906NA-I00.
\end{acknowledgements}

\bibliographystyle{aa}
\bibliography{bib} 
\begin{appendix}
\onecolumn
\section{Light curves}\label{alllc}

\begin{minipage}{\textwidth}
    \centering
    \includegraphics[width=0.45\textwidth]{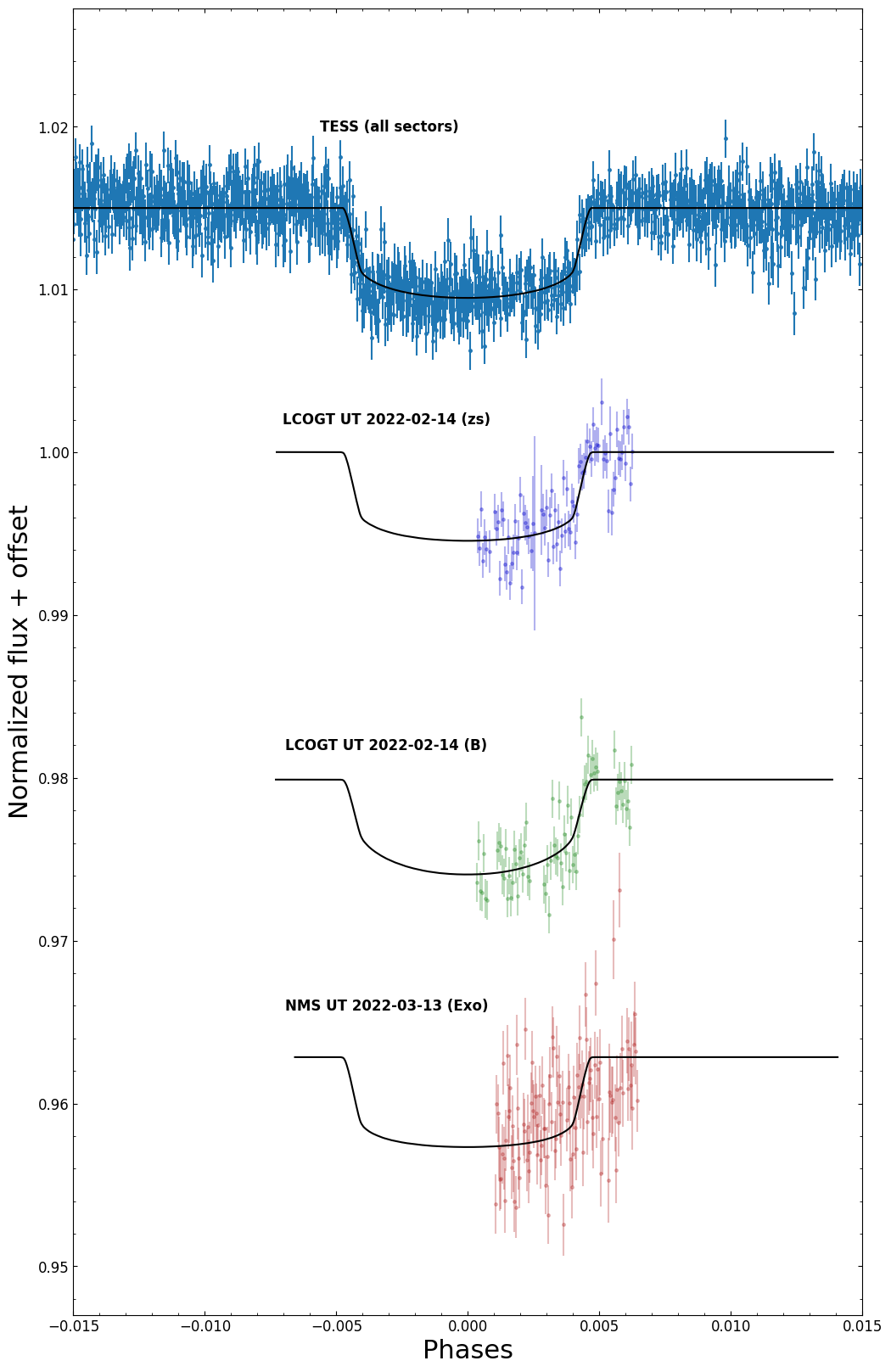}
    \includegraphics[width=0.45\textwidth]{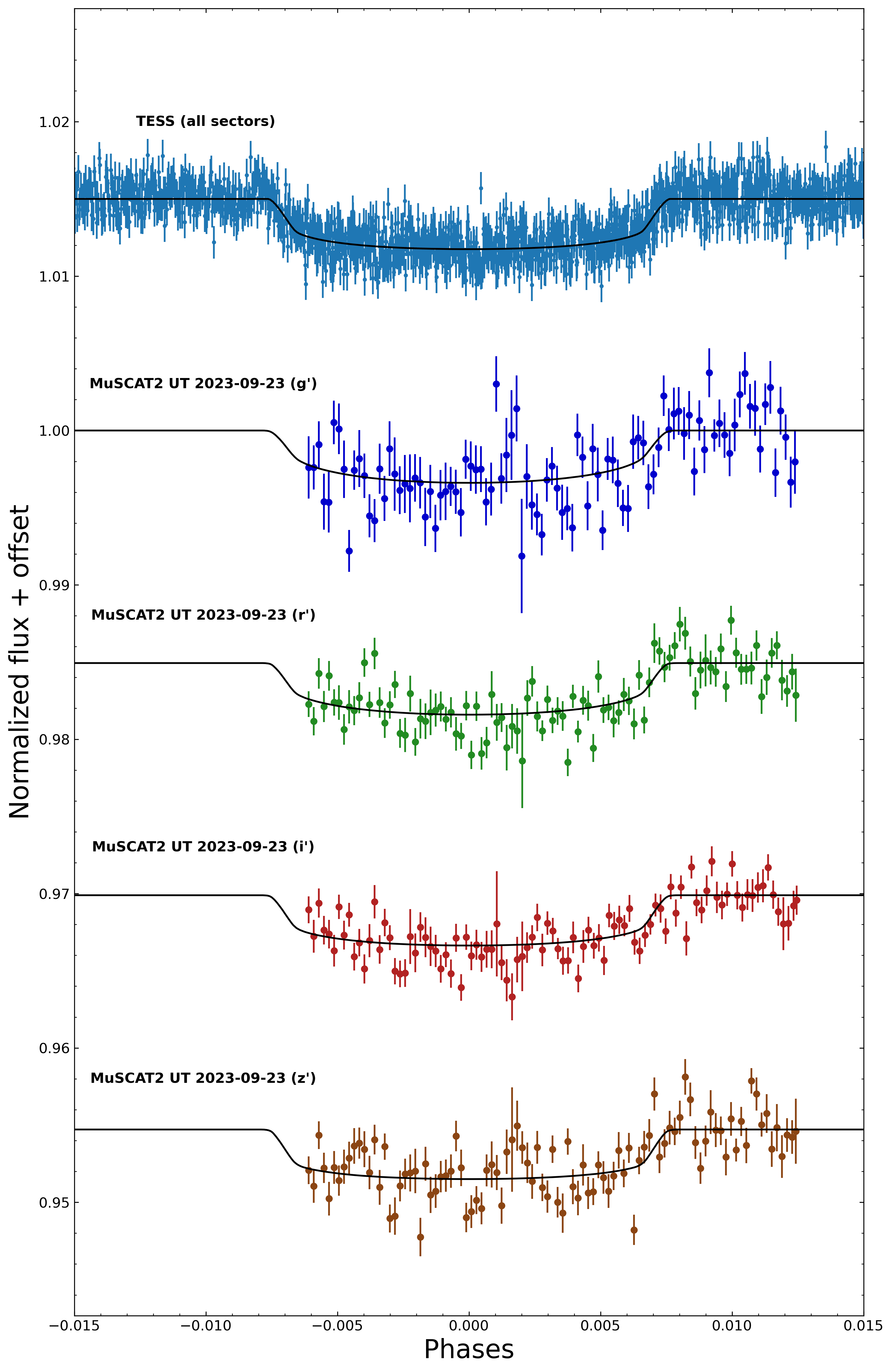}
    \captionof{figure}{Transit light curves of TOI-2147 and TOI-6019. The \textit{TESS} light curves are phase-folded with the data from all available sectors. The MuSCAT2 data for TOI-6019 is binned for better visibility.}
    \label{fig:lcall}
\end{minipage}
\FloatBarrier

\section{RV completeness plots} \label{kap:compl}
\begin{minipage}{\textwidth}
    \centering
    \includegraphics[width=0.45\textwidth]{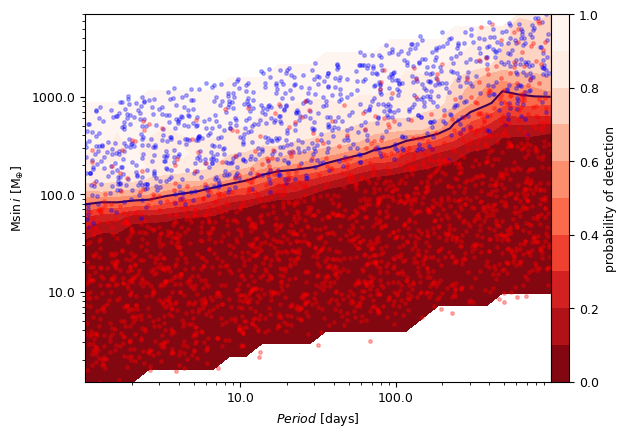}
    \includegraphics[width=0.45\textwidth]{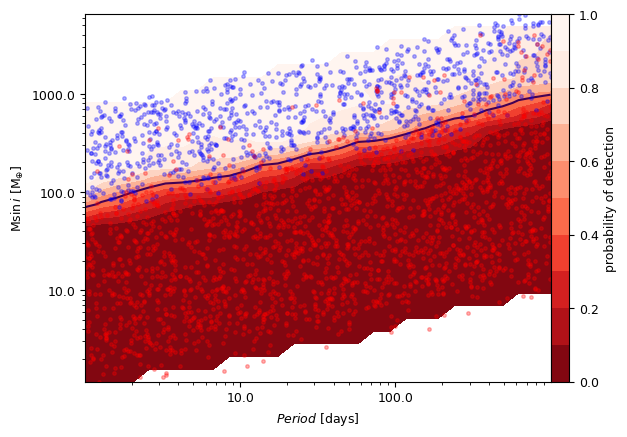}
    \captionof{figure}{Results of the injection-recovery tests for the RV data of TOI-2147 (left) and TOI-6019 (right). Blue points are injected planets that were recovered, while red points could not be detected. The shading of the regions represents the probability of detecting a potential companion in this region based on the data. For both stars, the data only allows to exclude the presence of massive nearby companions.}
    \label{fig:comp}
\end{minipage}
\FloatBarrier

\section{O-C diagrams} \label{kap:oc}
\begin{minipage}{\textwidth}
    \centering
    \includegraphics[width=0.49\textwidth]{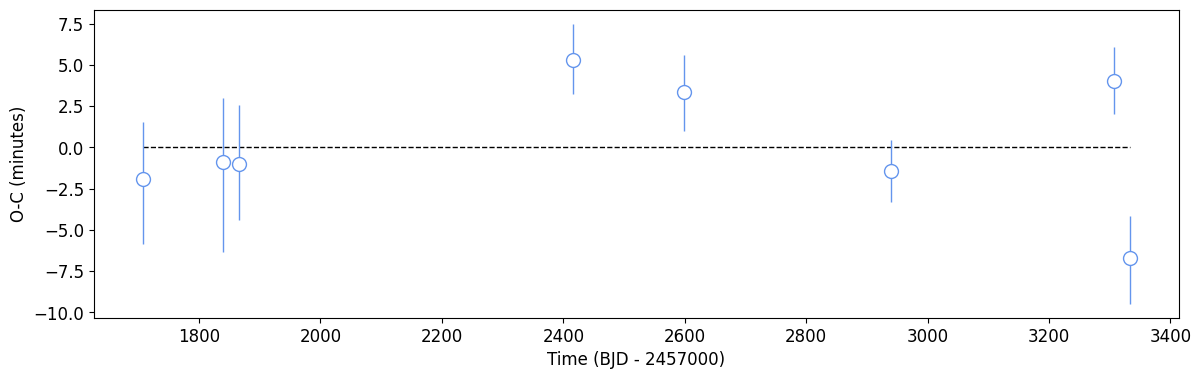}
    \includegraphics[width=0.49\textwidth]{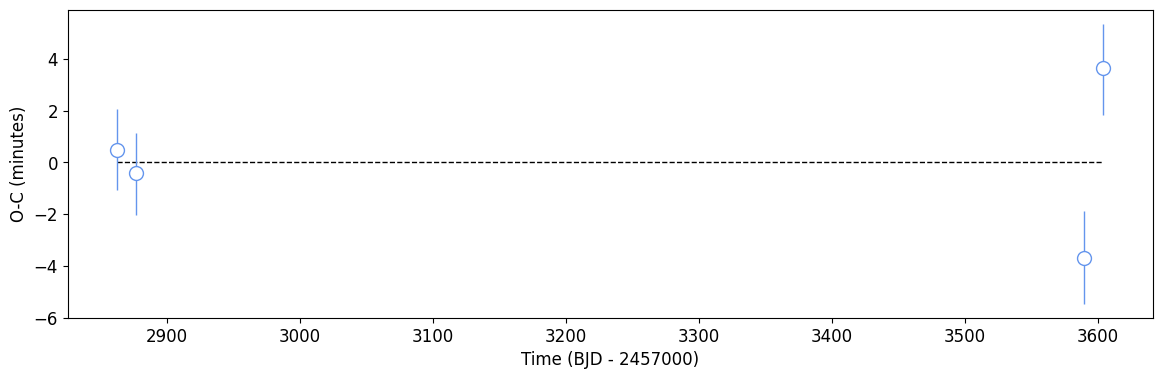}
    \captionof{figure}{O-C plots showing the measured TESS transit timing variations for TOI-2147\,b (upper plot) and TOI-6019\,b (lower plot).}
    \label{fig:oc}
\end{minipage}

\end{appendix}

\end{document}